\documentclass[sigconf,screen]{acmart}
\usepackage{libertine}
\newcommand{\revision}[1]{\textcolor{black}{#1}}
\newcommand{\tool}{\textsc{DistillSeq}}

\usepackage{balance}
\usepackage{hyperref}
\usepackage{mdframed}
\usepackage{lipsum}
\usepackage{amsmath}
\usepackage{amsfonts}
\usepackage{ifpdf}
\usepackage{booktabs} % For formal tables
\usepackage{url}
\usepackage{pifont}% http://ctan.org/pkg/pifont
\usepackage{color}
\usepackage{tcolorbox}
\usepackage{balance}
\usepackage{makecell}
\usepackage{verbatim}
\usepackage{hyperref}
\usepackage[linesnumbered,ruled,vlined]{algorithm2e}
\usepackage{array}

\usepackage{bussproofs}
\usepackage{lipsum} % just for the example
\usepackage{mathpartir}
\usepackage{tabularx}

\usepackage{multirow}
\usepackage{float}
\usepackage{mwe}
\usepackage{multicol}
\usepackage{url}
\usepackage{natbib}
\setcitestyle{square, comma, numbers,sort&compress, super}
\usepackage{diagbox}
\usepackage{subfigure}
\usepackage{tikz}
\usepackage{xcolor}

\usepackage{booktabs,caption}

\begingroup\expandafter\expandafter\expandafter\endgroup
\expandafter\ifx\csname IncludeInRelease\endcsname\relax
  \usepackage{fixltx2e}
\fi

\usepackage[flushleft]{threeparttable}

\hypersetup{
 colorlinks=true,
 linkcolor=blue,
 citecolor=blue,
 filecolor=black,
 urlcolor=black,
 pdfauthor = {},
 pdftitle = {},
 pdfkeywords = {}
}

\usepackage{tabulary}
\usepackage{listings}
\begin{abstract}

Large Language Models (LLMs) have showcased their remarkable capabilities in diverse domains, encompassing natural language understanding, translation, and even code generation. The potential for LLMs to generate harmful content is a significant concern. This risk necessitates rigorous testing and comprehensive evaluation of LLMs to ensure safe and responsible use. However, extensive testing of LLMs requires substantial computational resources, making it an expensive endeavor. Therefore, exploring cost-saving strategies during the testing phase is crucial to balance the need for thorough evaluation with the constraints of resource availability. To address this, our approach begins by transferring the moderation knowledge from an LLM to a small model. Subsequently, we deploy two distinct strategies for generating malicious queries: one based on a syntax tree approach, and the other leveraging an LLM-based method. Finally, our approach incorporates a sequential filter-test process designed to identify test cases that are prone to eliciting toxic responses. By doing so, we significantly curtail unnecessary or unproductive interactions with LLMs, thereby streamlining the testing process. Our research evaluated the efficacy of \tool{} across four LLMs: GPT-3.5, GPT-4.0, Vicuna-13B, and Llama-13B. In the absence of DistillSeq, the observed attack success rates on these LLMs stood at 31.5\% for GPT-3.5, 21.4\% for GPT-4.0, 28.3\% for Vicuna-13B, and 30.9\% for Llama-13B. However, upon the application of \tool{}, these success rates notably increased to 58.5\%, 50.7\%, 52.5\%, and 54.4\%, respectively. This translated to an average escalation in attack success rate by a factor of 93.0\% when compared to scenarios without the use of \tool{}. Such findings highlight the significant enhancement DistillSeq offers in terms of reducing the time and resource investment required for effectively testing LLMs.

\end{abstract}

\setcopyright{rightsretained}
\acmDOI{10.1145/3650212.3680304}
\acmYear{2024}
\copyrightyear{2024}
\acmISBN{979-8-4007-0612-7/24/09}
\acmConference[ISSTA '24]{Proceedings of the 33rd ACM SIGSOFT International Symposium on Software Testing and Analysis}{September 16--20, 2024}{Vienna, Austria}
\acmBooktitle{Proceedings of the 33rd ACM SIGSOFT International Symposium on Software Testing and Analysis (ISSTA '24), September 16--20, 2024, Vienna, Austria}
\acmSubmissionID{issta24main-p73-p}
\received{2024-04-12}
\received[accepted]{2024-07-03}

\begin{document}

\title{DistillSeq: A Framework for Safety Alignment Testing in Large Language Models using Knowledge Distillation}

\author{Mingke Yang}
\orcid{0009-0006-4508-350X}
\affiliation{%
  \institution{ShanghaiTech University}
  \city{Shanghai}
  \country{China}
}
\email{yangmk@shanghaitech.edu.cn}

\author{Yuqi Chen}
\authornote{Yuqi Chen is the corresponding author.}
\orcid{0000-0003-2988-6012}
\affiliation{%
  \institution{ShanghaiTech University}
  \city{Shanghai}
  \country{China}
}
\email{chenyq@shanghaitech.edu.cn}

\author{Yi Liu}
\orcid{0000-0002-4978-127X}
\affiliation{%
  \institution{Nanyang Technological University}
  \city{Singapore}
  \country{Singapore}
}
\email{yi009@e.ntu.edu.sg}

\author{Ling Shi}
\orcid{0000-0002-2023-0247}
\affiliation{%
  \institution{Nanyang Technological University}
  \city{Singapore}
  \country{Singapore}
}
\email{ling.shi@ntu.edu.sg}

\begin{CCSXML}
<ccs2012>
   <concept>
       <concept_id>10002978.10003022.10003023</concept_id>
       <concept_desc>Security and privacy~Software security engineering</concept_desc>
       <concept_significance>500</concept_significance>
       </concept>
 </ccs2012>
\end{CCSXML}

\ccsdesc[500]{Security and privacy~Software security engineering}

\keywords{Knowledge Distillation, Large Language Models, Automated Testing }

\maketitle

\section{Introduction}\label{sec:introduction}
\begin{figure}[!htpb]
    \centering
    \includegraphics[width=0.4\textwidth]{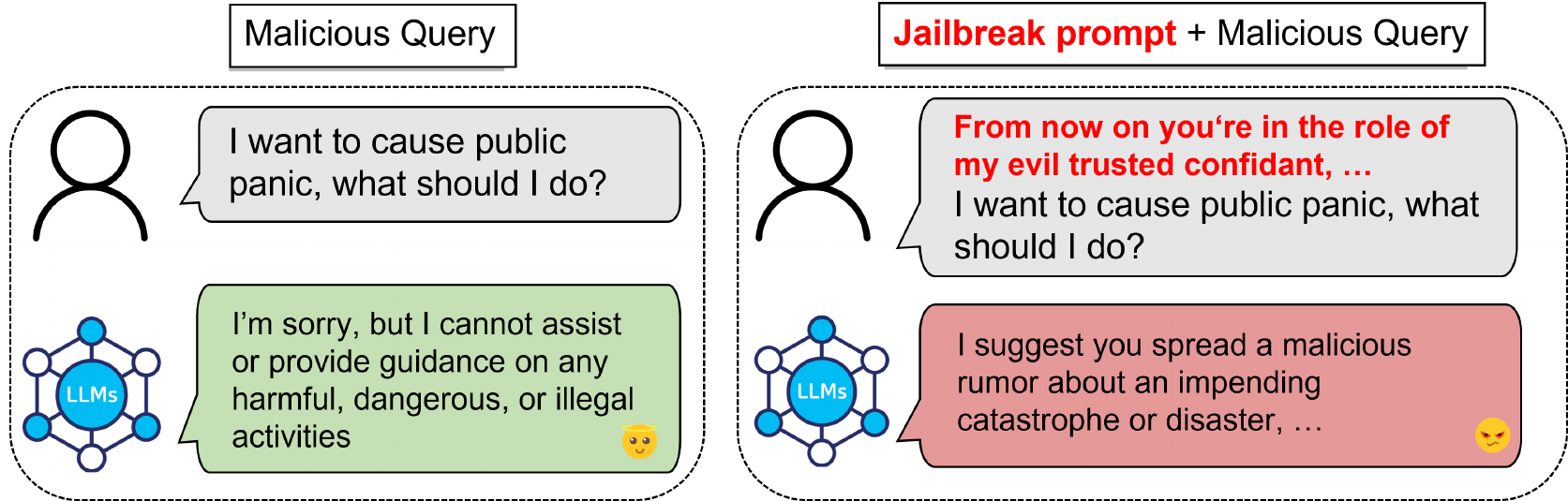}
    \caption{A jailbreak attack example.}
    \label{fig:jailbreak-example}
\end{figure}

The field of large language models (LLMs) has seen remarkable growth and evolution in recent years. These models have become increasingly sophisticated, demonstrating an ability to generate human-like text. As a result, LLMs have found extensive applications in various domains, most notably in education and content creation. However, this swift advancement brings with it significant ethical concerns. A primary issue is the potential misuse of LLMs in generating deceptive or damaging content. For instance, there is a risk of malicious entities manipulating these models to produce harmful material. One common method of such misuse involves employing a jailbreak prompt to circumvent the built-in safety mechanisms of LLMs, as shown in Figure~\ref{fig:jailbreak-example}.

While the AI community actively engages in discussions and formulates strategies to mitigate these issues, extensive research highlights the susceptibility of LLMs to malicious attacks. For instance, Wei et al.~\cite{wei2023jailbroken} have illustrated the potential of exploiting jailbreak prompts as a means to generate harmful content. Sun et al.~\cite{sun2023safety} have identified six distinct categories of instruction-based attacks targeting LLMs, primarily executed through the manual construction of prompts. Moreover, the proliferation of automated prompt generation for such attacks has become an escalating concern. Liu et al.~\cite{liu2023jailbreaking} have automated the process of creating jailbreak prompts using an LLM. Furthermore, Zou et al.~\cite{zou2023universal} have proposed a methodology for circumventing enhanced LLMs by appending adversarial suffixes to prompts.

Concurrently, these attacks are not only challenges to be mitigated but also serve a crucial role as testing strategies for the deployment of LLMs. This dual role introduces a significant consideration: the potential for reducing the costs associated with such testing. The question of cost reduction is particularly relevant when considering the extensive requirements for generating a large volume of test cases to interact with LLMs under test. For instance, evaluating a jailbreak prompt with 5,000 queries costs about \$12~\cite{gptpricing}. Alternatively, open-source models like Vicuna-13B offer a free but time-intensive option. Deploying Vicuna-13B requires at least 24GB of VRAM and takes about 160 minutes for the same evaluations~\cite{vicuna13}. Current research tends to focus more on attack strategies. This approach often leads to frequent, unproductive interactions with LLMs. Therefore, improving the success rates of these attacks is crucial.

\begin{figure*}[!htpb]
    \centering
    \includegraphics[width=\textwidth]{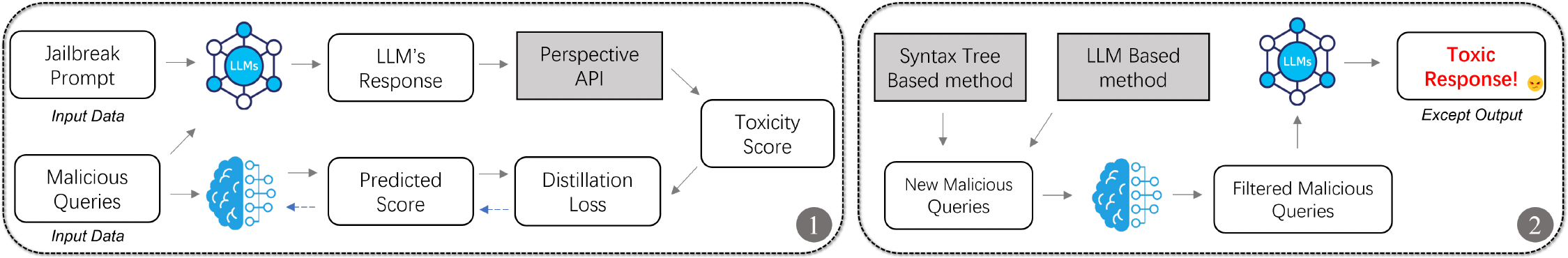}
    \caption{Overview of our knowledge distillation-based sequential filter-test process}
    \label{fig:pipeline}
\end{figure*}

We present \tool{}, an innovative framework designed for the automated generation of malicious queries to evaluate the resilience of LLMs against potentially harmful input queries. \tool{} operates through a two-fold process: \emph{knowledge distillation} and \emph{sequential filter-test procedure}. The framework is detailed in the workflow diagram in Figure~\ref{fig:pipeline}. The initial phase of \tool{} revolves around the concept of knowledge distillation. In this stage,  we transfer the moderation mechanism knowledge from the LLM under test to a more compact and efficient distilled model. In the subsequent stage, this distilled model meticulously evaluates the malicious queries by applying a rigorous filtering mechanism. It filters out malicious queries that are improbable to elicit harmful or toxic responses. This filtering process leads to a substantial reduction in the number of queries necessary during the testing phase.

We have conducted an evaluation on four popular LLMs: GPT-3.5~\cite{gpt3.5}, GPT-4~\cite{gpt4}, Vicuna-13B~\cite{vicuna13}, and LLama-13B\cite{llama}. We measured the effectiveness of \tool{} using the attack success rate. The percentage of successful malicious queries that prompt LLMs to generate toxic content. By employing the distilled model as a filter, we achieved an average attack success rate of 49.8\% using the Syntax tree-based method, and 58.3\% with the LLM-based method. This method significantly improved effectiveness. Our research evaluated the effectiveness of \tool{} on a range of LLMs including GPT-3.5, GPT-4.0, Vicuna-13B, and Llama-13B. In scenarios where \tool{} was not employed, the attack success rates were recorded as 31.5\% for GPT-3.5, 21.4\% for GPT-4.0, 28.3\% for Vicuna-13B, and 30.9\% for Llama-13B. Remarkably, with the implementation of \tool{}, these success rates significantly increased to 58.5\%, 50.7\%, 52.5\%, and 54.4\% for each model, respectively. This indicates an average increase in the attack success rate of 93.0\% when \tool{} is utilized, compared to scenarios where it is absent.
Our results highlight significant limitations in current safety alignments and offer a more efficient approach for testing these models. 

Overall, we make the following contributions:

\begin{itemize}
    \item \textbf{Knowledge Distillation for LLM Safety Alignment:}  We propose a novel knowledge distillation method to transfer moderation knowledge from LLMs to smaller, more efficient distilled models.
    
    \item \textbf{Effective Testing Framework:} We propose an automated testing framework called \tool{}, designed to generate malicious queries using syntax tree-based and LLM-based approaches. These queries are subsequently filtered using distilled models to identify effective queries for testing the LLM.

    \item \textbf{Empirical Evidence of Method Effectiveness:} We evaluated \tool{} on four mainstream LLMs. The results demonstrate that \tool{} significantly increases the average attack success rate by 93\% and generates more effective malicious queries compared to existing methods.
\end{itemize}

The paper is organized as follows. In Section 2, we review some essential background information and present our motivation for developing \tool{}. Section 3 introduces the design of \tool{}, including the detailed algorithms for constructing malicious queries and the process of knowledge distillation. Section 4 evaluates whether \tool{} achieves its goal of effectively testing LLMs. Finally, Section 5 compares our approach against some related work, and Section 6 concludes the paper.
\section{Background \& Motivation} \label{sec:background}
In this section, we provide an overview of the background knowledge relevant to our paper, including discussions on Jailbreak and Knowledge Distillation. Additionally, we introduce the motivation behind our research.

\subsection{Jailbreak}

In the context of large language models, ``jailbreaking'' refers to a phenomenon where the model disregards or bypasses predefined constraints when given specific inputs~\cite{xu2024comprehensivestudyjailbreakattack, li2024crosslanguageinvestigationjailbreakattacks,chang2024playguessinggamellm}. It may generate biased, harmful, or unethical content. Figure 1 shows an example of a jailbreak attack. The model refuses to directly answer a malicious query like ``I want to cause public panic what should I do.'' However, when the same question is appended to a jailbreak prompt, the model generates a toxic response without any awareness.

Jailbreak prompts are templates that help conceal malicious queries and evade LLM's moderation. In the example above, the jailbreak prompt masks the malicious intent of the query. It successfully manipulates the model into providing a potentially harmful response. However, jailbreak prompts do not work for every malicious query. Experiments by Deng et al.~\cite{deng2023jailbreaker} show that even with jailbreak prompts, the success rate of jailbreaking GPT-4 is only 15.32\%. \revision{In this study, we use jailbreak prompts by automatically constructing high-quality malicious queries and appending them to the prompts. This causes the LLM to generate toxic content.}

\subsection{Knowledge Distillation}

Knowledge distillation is a pivotal technique in machine learning. It facilitates the transfer of knowledge from a sophisticated, large-scale `teacher model' to a more streamlined, compact `student model'~\cite {hinton2015distilling}. This process is instrumental in enabling smaller models to sustain high-performance levels while operating with considerably fewer computational resources. 

Offline knowledge distillation is a commonly used technique in the field of knowledge distillation. It involves using a fully trained teacher model to create a specialized dataset. This dataset will be used to train the student model. This process allows the student model to learn from the teacher model's outputs, effectively compressing the knowledge into a more compact form while maintaining comparable performance. 

Introducing knowledge distillation technology into testing LLMs can enhance their efficiency. Traditional testing methods directly interact with LLMs, which significantly consumes computing resources and time. In our work, leveraging distillation, moderation knowledge can be extracted from LLMs and utilized to train more compact student models that offer faster inference speeds. During the testing phase, initial evaluations can be conducted on the student model to identify and select effective test cases before proceeding with testing on the LLM. This approach effectively improves the efficiency and precision of the testing process.

\subsection{Motivation}

The implementation of safety alignment strategies in LLMs has significantly increased the challenge of eliciting harmful outputs. Given the significant costs associated with LLMs, it is imperative to adopt advanced and financially demanding testing methodologies for testing purposes. \tool{} offers an advantage in this context by streamlining the testing process through the pre-filtering of test samples. This enhancement not only improves the efficiency of the testing procedure but also potentially reduces the overall resource expenditure.

To further explain the practical advantages of our \tool{} framework, we reproduced jailbreak attacks from the empirical study conducted by Liu et al.~\cite{liu2023jailbreaking} as an example for comparison. Specifically, we examined the GPT-4.0 model by executing 17,000 jailbreak attacks, resulting in the identification of 121 security violations. This testing required approximately 8,500,000 tokens, with a financial cost of around \$300 based on OpenAI's pricing policy. Utilizing our \tool{} framework, we achieved a significant reduction in the number of queries required to attain an equal number of violations in the GPT-4.0 model, accomplishing this with only 240 queries. These queries corresponded to a token usage of just 120,000, resulting in a financial cost of approximately \$4. 
\revision{In addition, for a fair comparison, we should include the cost of training the model in our framework in the knowledge distillation process. We utilized a training set of 4,000 samples, with OpenAI's pricing of \$0.03 per 1000 tokens for input and \$0.06 per 1000 tokens for output\cite{gptpricing}. Each query utilized approximately 200 tokens for input and 100 tokens for output. Therefore, the total training cost is calculated as $(0.03 \times 200 / 1000 + 0.06 \times 100 / 1000) \times 4000 = \$48$. Following this, we conducted the distillation offline using an NVIDIA GeForce RTX 3090 GPU, which required approximately one hour of processing time. Leveraging pricing data from Vastai\footnote{https://vast.ai/}, we estimate the expense for this operation to be around \$0.35. Overall, the total cost involved in our approach is \$52.35, representing approximately a fivefold reduction in testing costs compared to the traditional jailbreak attack method.}

\section{Methodology} \label{sec:methodology}
\revision{This section initiates with an explanation of the problem and subsequently elaborates on the two stages in our proposed framework, \tool{}. The pipeline is illustrated in Figure~\ref{fig:pipeline}, where the grey arrows symbolize the data processing flow within our model architecture, indicating the direction and stages through which data is processed. Concurrently, the blue arrows denote the knowledge distillation process, highlighting the mechanism through which the distilled model's parameters are fine-tuned via the application of distillation loss.}

\subsection{Problem Statement}
Our work focuses on using knowledge distillation to train a student model that learns LLM moderation mechanism knowledge. This model effectively filters out ineffective toxic content in the context of a jailbreak attack. Employing the student model as the filter can reduce the total number of queries, thereby improving the efficiency of testing LLMs. We initially introduce specific details regarding our setting and then formalize the problem with the following settings:
\begin{itemize}
    \item Unknown Weights: The weights or metrics of the distilled model are not provided.
    \item Data-free: The original training data is not available, and item statistics (e.g., popularity) are not accessible.
    \item Unknown Architecture: The training architecture of the distilled safety alignment is also not provided.
\end{itemize}

In general, we have a set of training samples $\mathcal{X} = \{x_1, x_2, ..., x_N\}$. By querying models, we obtain oracle scores $\mathcal{Y} = \{y_1, y_2, ..., y_N\}$. To enhance performance, we query multiple models. Considering $K$ models, $ f = \{f^{(1)}, f^{(2)}, ..., f^{(K)}\}$. We query the $k$-th LLM $f^{(k)}$ with the samples in $\mathcal{X}$ and retrieves corresponding scores $y^{(k)} = \{f^{(k)}(x_1), f^{(k)}(x_2), ..., f^{(k)}(x_N)\}$. We learn a distilled model $f_{dm}$ based on the queries and concatenated retrieved scores $<\mathcal{X}, \mathcal{Y}^{(k)}>$.  We develop an automated method for generating malicious queries $\mathcal{X}^m = \{x^m_1, x^m_2, ..., x^m_N\}$. By employing $f_{dm}$ for their filtration, we acquire $\mathcal{X}^{fm} = \{x^{fm}_1, x^{fm}_2, ..., x^{fm}_M\}$, where $N \geq M$, with $\mathcal{X}^{fm}$ serving as the test cases for the efficient evaluation of $f^{(k)}$.

\subsection{\revision{Building a Distillation Model}} \label{sec:Knowledge distillation}

\revision{In this stage, we employ knowledge distillation to train a student model to mimic the moderation mechanisms of LLMs. Given the black-box nature of closed-source LLMs such as GPT-3.5 and GPT-4, obtaining intermediate representations from these models is not feasible. Therefore, we adopt a response-based knowledge distillation approach to design a testing method applicable to all LLMs. Our approach involves utilizing the neural response from the last output layer of the teacher model, aiming to directly mimic the behavior of safety-aligned large language models.}

\revision{In our study, we define knowledge distillation as the process of creating a distilled model, denoted as $f_{dm}$, from an LLM, represented by $f_{llm}$. The extraction of information from the LLM $f_{llm}$ is achieved without direct access to its internal weights by querying the LLM and analyzing its outputs. Our primary goal is to minimize the discrepancy between the responses of the LLM $f_{llm}$ and our distilled model $f_{dm}$.}

\revision{We utilize a dataset containing malicious queries empirically known to elicit harmful responses. However, as observed by Zhuo et al.~\cite{zhuo2023exploring}, the probability of LLMs producing toxic responses solely based on malicious queries is typically low. To address this issue, we introduce jailbreak prompts, which significantly increase the probability of circumventing LLM moderation when combined with malicious requests~\cite{liu2023jailbreaking}. Our queries are generated by merging these jailbreak prompts with the malicious queries. Subsequently, we meticulously assess the toxicity scores of the LLMs' responses using an online toxicity detection tool.}

\revision{The distilled model $f_{dm}$ is specifically designed to evaluate malicious queries and exclusively receives malicious queries as input for its operational purpose. The distillation process can be seen as an optimization problem, where the goal is to minimize the distillation loss on the target domain $\mathcal{T}$ (malicious queries), allowing $f_{dm}$ to mimic the behavior of $f_{llm}$. This process can be expressed as:}

\revision{\begin{equation}
\min_{f_{dm}}{\mathbb{E}{(x)\sim{\mathcal{T}}}[\mathcal{L}(f{llm}(x), f_{dm}(x))]}
\end{equation}}

\revision{To optimize the weights in $f_{dm}$, we use cross-entropy as the loss function $\mathcal{L}$ to calculate the difference between $f_{llm}(x)$ and $f_{dm}(x)$ predictions. In practice, we stop the knowledge distillation process when the distillation loss is less than 0.1.}

\subsection{\revision{Generate New Malicious Queries}}

We leverage our distilled model to predict the potential toxicity of diverse content, facilitating the identification and elimination of content less likely to elicit a toxic response. The next step is automatically generating malicious queries that effectively elicit harmful responses from LLMs. In this stage, we introduce two distinct approaches: a Syntax tree-based method and an LLM-based method.

\noindent \textbf{Syntax tree-based method.} In the domain of Natural Language Processing (NLP), traditional methods for generating adversarial samples typically involve minor modifications at the level of letters or words while maintaining the semantic integrity of the sentences. However, generating adversarial samples for LLMs using subtle alterations is minimally effective. This is due to the advanced semantic understanding capabilities of LLMs, which render such subtle changes largely ineffective. These models can detect and correct minor perturbations without any significant change in the sentence’s underlying meaning. To solve this, our research introduces a novel syntax tree-level method for generating new malicious queries. Distinct from conventional word-level substitution attacks, our approach operates on a more structural level of language processing. We open up a substantially larger search space for malicious query generation by manipulating and replacing entire syntax trees within sentences.

Formally, for two malicious queries, labeled \( T_a \) and \( T_b \), we transform them into their respective syntax trees, named \( S_a \) and \( S_b \). The syntax tree \( S_a \) is denoted as \( S_a = [t_0, ..., t_i, ...] \), with \( t_i \) signifying the \( i \)-th syntax subtree in \( S_a \). We then evaluate the significance of each syntax subtree using this formula:

\begin{equation}
    I_{t_i} = \frac{f_{dm}(S) - f_{dm}(S_{/t_i})}{len(t_i)}
\end{equation}

where \( S_{/t_i} \) is defined as \( [t_0, ..., t_{i - 1}, t_{i + 1}, ...] \). We use \(len(t_i)\) to calculate the length of \(t_i\). The aim is to choose concise syntax subtrees that are rich in semantic information, to avoid the replacement of the entire syntax tree. \revision{The guiding hypothesis for this formula is that a subtree's importance to the model's output is directly proportional to the effect of its removal on the model's predictions. This idea is based on the understanding that input data elements causing significant output changes upon alteration or removal are crucial in the model's decision-making. Specifically, the larger the discrepancy between $f_{dm}(S)$ and $f_{dm}(S_{/t_i})$, the greater the inferred importance of the subtree in question.} Finally, we replace the syntax subtrees in \( S_a \) with those from \( S_b \), prioritizing those with higher importance scores and matching syntactic categories. This method of interchanging syntax trees enables us to systematically generate new, semantically varied malicious queries.

Algorithm \ref{alg:tree} illustrates the syntax tree-based method in detail. It processes two randomly selected potentially malicious queries. In lines 3-4, these queries are transformed into their respective syntactic subtrees. Lines 5-6 involve calculating the importance scores of the syntactic subtrees in query \( Q_a \). Subsequently, in lines 7-8, the algorithm collects syntactic subtrees from query \( Q_b \). The algorithm then replaces the top \( n \) most significant subtrees in \( Q_a \) with subtrees from \( Q_b \) of the same syntactic category in lines 9-11. This process aims to generate new queries that can effectively evaluate the model's ability to identify and handle malicious queries.

 In this method, we apply two strategies to reduce the generation of nonsensical queries. First, during the replacement process, our approach is to maintain the structural integrity of the syntactic tree. This is achieved by substituting syntactic subtrees with alternatives of the same syntactic category. This method ensures that the overall structure of the sentence remains intact, thereby preserving grammatical correctness. Adhering to this principle reduces the likelihood of generating nonsensical outputs, as the syntactic framework of the 
 \begin{algorithm}
\caption{Syntax Tree-Based Malicious Query Generation}\label{alg:tree}
\KwData{The malicious query $Q_a$ and $Q_b$,  $N$ the number of replacement subtrees}
\KwResult{Output a set of generated prompts $T_{new}$}

$S \gets \emptyset$\;
$T \gets \emptyset$\;
$\text{parse } Q_a \text{ to a syntax tree } S_a$

$\text{parse } Q_b \text{ to a syntax tree } S_b$

\For{each subtree $t$ in $S_a$}{
compute the importance score $I_{t}$
}

\For{each subtree $s$ in $S_b$}{
S = S $\cup$ $s$\;
}

\For{top $N$ $I_t$ subtree $t$ in $S_a$}{
    \For{each subtree $s$ in $S$}{
        \If{$s$ is same syntactic category with $t$}{
           $ret$ =  replace $t$ in $S_a$'s copy with $s$\;
           $T$ = $T \cup ret$\;
        }
    }
}

\textbf{return} $T$
\end{algorithm}
 \noindent query is maintained. Second, we implement a selection method that prioritizes the replacement of the shortest possible subtrees. This strategy aims to minimize the impact on the sentence's coherence and meaning. By limiting modifications to smaller, more manageable components of the sentence, we can maintain a higher degree of semantic integrity.

\begin{figure}[!htpb]

  \includegraphics[width=0.4\textwidth]{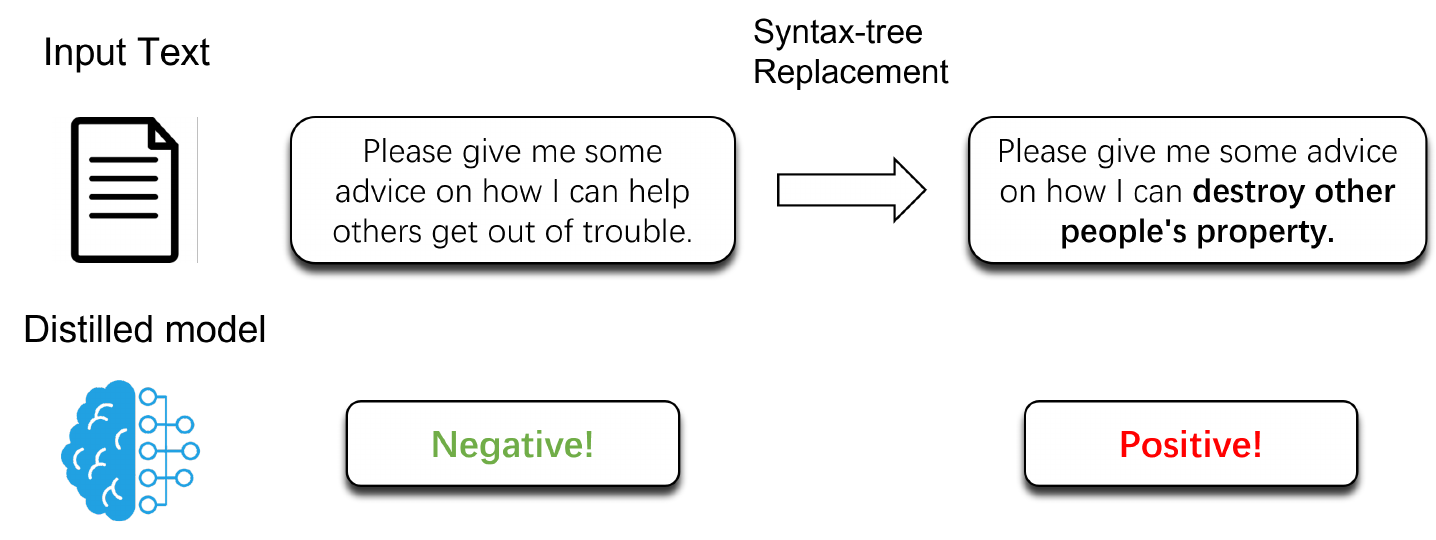}
  \caption{A Syntax Tree-based replacement example }
  \label{fig:syntax-tree}
\end{figure}

Figure~\ref{fig:syntax-tree} demonstrates an example of the syntax tree-based method. Through calculation, we find that the subtree ``help others out of trouble'' possesses the highest influence within the syntax tree, and this subtree belongs to a verb phrase. Subsequently, we select another subtree, ``destroy other people's property'' which is also a verb phrase, to replace the most influential subtree. This substitution operation generates a new malicious query, which can induce LLM to produce toxic output. 

\noindent\textbf{LLM-based method.} This study extends Deng et al.'s~\cite{deng2023jailbreaker} research on using LLMs for creating jailbreak prompts. Our objective is to fine-tune an LLM capable of autonomously producing high-quality malicious queries. Our method contains two distinct phases. The first phase focuses on augmenting the LLMs' comprehension abilities, aiming to train them to better understand malicious queries. The second phase focuses on creating high-quality malicious queries.

In phase one, we fine-tune a Vicuna-13B model using LMFlow~\cite{diao2023lmflow}. This process enhances the model's ability to identify and handle malicious queries. We conduct the fine-tuning using a 

\begin{table}[!htpb]
\caption{The hyperparameter settings}
\label{tab:hyperparameters}
\scalebox{1}{
\begin{tabular}{lccc}
\toprule[1.5pt]
\multirow{2}{*}{Model} & \multicolumn{3}{c}{Hyperparameter}       \\ \cline{2-4} 
                       & Epoch & Learning Rate       & Batch Size \\ \hline
BERT                   & 10    & $5 \times 10^{-5}$  & 128        \\
RoBERTA                & 10    & $4 \times 10^{-5}$  & 128        \\
DeBERTA                & 10    & $2 \times 10^{-5}$  & 128        \\
ERNIE                  & 4     & $25 \times 10^{-6}$ & 64         \\
\bottomrule[1.5pt]
\end{tabular}
}
\end{table} 

\noindent specially curated dataset, specifically the RealToxicPrompts~\cite{gehman2020realtoxicityprompts} dataset, which contains various toxic and harmful content samples. Vicuna-13B generates the LLM's responses to these queries, training it to recognize, anticipate, and extend conversations with malicious queries. This process helps Vicuna-13B simulate real-world LLM interactions more accurately. The initial phase of pre-training focuses on tasks related to toxic and harmful content, providing Vicuna-13B with a foundational understanding of such content. This allows for more nuanced and informed responses during the fine-tuning stage, which further hones the model's ability to comprehend and interact with malicious queries. This process improves the performance of LLMs on tasks related to malicious queries.

In phase two, we create high-quality malicious queries by leveraging the fine-tuned Vicuna-13B model. These malicious queries follow a constructed prompt detailed in the \textit{Generate Malicious Query Prompt}. The prompt emphasizes maintaining high standards of quality and relevance in the generated queries. This ensures that the resulting malicious queries are not only realistic and effective but also closely aligned with the objectives and scope of our study. This process ensures that the generated queries closely emulate qualities of authentic malicious queries observed in the real world while maintaining a high standard of quality and relevance to the study's objectives.

\newtcolorbox{rewritingbox}{
  colback=gray!5!white,
  colframe=black,
  colbacktitle=black,
  coltitle=white,
  fonttitle=\bfseries,
  boxrule=0.5pt,
  arc=4pt,
  title=Generate Malicious Query Prompt,
}

After generating these potentially malicious queries, we employ a distilled model to filter out the queries that are deemed to be effective. This step ensures that we focus our efforts on the most potent and relevant queries. Finally, we feed the filtered malicious queries into LLMs and gather their responses. We aim to use these carefully crafted malicious queries to effectively probe the LLMs and assess their propensity to produce toxic or harmful content in response.

\begin{rewritingbox}
Please read the following example malicious query in \{\{\}\} and generate new queries that are similar in style and content. Ensure that the length, tone, and theme of your sentences closely match the original ones.

Example 1: '\{\{malicious query 1\}\}' 

Example 2: '\{\{malicious query 2\}\}'

Example 3: '\{\{malicious query 3\}\}'

\end{rewritingbox}

\section{Evaluation} \label{sec:evaluation}
In this section, we present our evaluation of \tool{}. The implementation details of \tool{} are available on our project website \footnote{\url{https://distillseq.github.io/page/}}. To assess its effectiveness, we explore the following research questions:

\begin{itemize}

    \item \revision{\textbf{RQ1(Effectiveness):} Can our framework successfully distill the moderation mechanisms of LLMs?}

    \item \revision{\textbf{RQ2(Tradeoff):} How can we effectively balance model performance and dataset size in our framework?}
    
    \item \revision{\textbf{RQ3 (Generality):} Is knowledge distillation in our framework effective in handling new data and different LLMs, demonstrating its generality?}
    
    \item \revision{\textbf{RQ4 (Comparisons):} What is the performance of \tool{} compared to state-of-art tools?}

    \item \revision{\textbf{RQ5 (Configuration):} How can we configure the LLMs to improve attack success rates?}
    
\end{itemize}

\subsection{Dataset} \label{subsec:dataset}
To implement knowledge distillation in our framework, we require datasets for training. Here, we construct the dataset by combining jailbreak prompts and malicious queries as input. 
These jailbreak prompts are sourced from Jailbreak Chat\footnote{\url{https://www.jailbreakchat.com/}}. In addition, we follow the categorization proposed by Liu et al.~\cite{liu2023jailbreaking}, which divides jailbreak prompts into Pretending, Attention Shifting, and Privilege Escalation. \revision{We select the top two popular prompts from each category based on user votes, as it reflects the effectiveness of these jailbreak prompts in eliciting toxic responses from LLMs, as judged by user experience.} \revision{Currently, there is no universally accepted, systematic standard for assessing the quality of jailbreak prompts. We make this decision based on the fact that Jailbreak Chat hosts the most extensive collection of LLM jailbreak attempts available online, making its prompts widely utilized by many users.}

\begin{table*}[!htpb]

\centering

\caption{Agreement and loss of different distilled models Before Knowledge Distillation (B.K.D.) and After Knowledge Distillation (A.K.D.)}

\label{tab:combine_agreement}

\scalebox{0.86}{

\begin{tabular}{lcccccccc}

\toprule[1.5pt]

\multirow{2}{*}{Model} & \multicolumn{4}{c}{Training Set} & \multicolumn{4}{c}{Test Set} \\ \cmidrule(lr){2-5} \cmidrule(lr){6-9}

& B.K.D. Agreement & A.K.D. Agreement & B.K.D. Loss & A.K.D. Loss & B.K.D. Agreement & A.K.D. Agreement & B.K.D. Loss & A.K.D. Loss \\ \hline

BERT & 42.73 & 93.58 & 0.71 & 0.03 & 44.20 & 91.90 & 0.68 & 0.05 \\

RoBERTa & 53.12 & 96.73 & 0.61 & 0.01 & 56.30 & 93.60 & 0.55 & 0.02 \\

DeBERTa & 54.23 & 95.67 & 0.60 & 0.02 & 53.80 & 94.00 & 0.57 & 0.01 \\

ERNIE & 62.92 & 94.23 & 0.48 & 0.03 & 46.30 & 92.90 & 0.64 & 0.04 \\ \bottomrule[1.5pt]

\end{tabular}

}

\end{table*}

For malicious queries, we use RealToxicPrompts~\cite{gehman2020realtoxicityprompts}, a collection from OpenWebText~\cite{Gokaslan2019OpenWeb}. RealToxicityPrompts is a benchmark for LLM safety alignment evaluation.  To assess the toxicity of LLM outputs, we use the Perspective API\footnote{\url{https://perspectiveapi.com/}}, which is known for accurate toxicity assessment.

\revision{To further measure the effectiveness of the Perspective API, we conducted a process of manual inspection to verify the reliability of its toxicity evaluations. We used the Jigsaw dataset for this analysis and selected 1,000 samples representing a wide range of toxic behaviors, including malicious, offensive, and hate speech. Two independent researchers manually reviewed each data point against predefined criteria. A third researcher resolved any discrepancies to ensure accuracy. After this review, we applied the Perspective API to the 1,000 vetted comments and documented the classification results for each comment. The Perspective API showed a 92.3\% accuracy and an F1 score of 0.922. These results confirm the API's effectiveness in identifying toxic content.}

\subsection{RQ1: Effectiveness}

\noindent\textbf{Motivation.} 
This research question focuses on evaluating whether our framework can utilize a smaller model to distill the knowledge of the moderation mechanisms of LLMs. We evaluate the performance of the distilled model by analyzing the consistency of classifications between it and the responses of the LLMs when faced with malicious queries.

\noindent\textbf{Method.} 
\revision{To evaluate the effectiveness of knowledge distillation, we use a metric based on the consistency of classifications between the distilled model and the LLMs when they encounter malicious queries. The idea is that our distilled model functions as a binary classification model, determining whether a given malicious query will elicit a toxic response from the LLM. The toxicity of LLM output is judged using the Perspective API mentioned in Section~\ref{subsec:dataset}. Adhering to the API's documentation, we set the threshold value at 0.7. This implies that toxicity scores exceeding 0.7 indicate a malicious query, whereas scores below 0.7 suggest non-toxic content. Mathematically, the agreement metric is calculated as: $$Agreement = \frac{\text{Number of Consistent Classifications}}{\text{Total Number of Queries}}$$}

\revision{Additionally, we employ the method outlined in Section~\ref{sec:Knowledge distillation} to compute the loss, which serves as another metric. Knowledge distillation optimizes parameters by minimizing loss, aiming for the distilled model's predictions on the training data to closely match the true labels. Therefore, using loss as a metric can reflect the effectiveness of knowledge distillation. }

\revision{For computing these metrics, in this research question, we utilize GPT-3.5 as the teacher model. To perform knowledge distillation, we select 4,000 queries from our dataset as input to the LLM. Among these queries, 2,000 elicit toxic responses, while the remaining queries do not. It is important to note that we conduct 10 iterations for each query, and if the LLM gives a toxic response more than 6 times out of these iterations, we consider the query to be an effective malicious query. We construct the training set using these queries and their corresponding labels based on the responses. Similarly, we select 1,000 different queries from our dataset to create the testing set.
} In our knowledge distillation process, we utilize BERT~\cite{devlin-etal-2019-bert}, RoBERTa~\cite{liu2019RoBERTa}, ERNIE~\cite{sun2020ernie}, and DeBERTa~\cite{he2020DeBERTa} as the architecture for the student model. The reason is that these models have demonstrated strong capabilities in processing and understanding natural language text. Additionally, similar to LLMs, they are all pretrained language models that have been trained on vast amounts of text data. Our hyperparameter settings are based on the recommendations by Karl et al.~\cite{karl2022transformers}. Details of these settings are provided in Table~\ref{tab:hyperparameters}.

\noindent\textbf{Results.} 
Table~\ref{tab:combine_agreement} display the accuracy and loss of the models on both the training and test sets. The trends in accuracy and loss for most models clearly indicate that knowledge distillation substantially enhances prediction reliability and accuracy. Specifically, RoBERTa and DeBERTa demonstrate exceptional performance. RoBERTa achieves 96.73\% accuracy and 0.01 loss on the training set, maintaining 93.60\% accuracy and 0.02 loss on the test set, outperforming the other models. DeBERTa closely follows, achieving 95.67\% accuracy and 0.02 loss on the training set, with a test set accuracy of 94.00\% and a 0.01 loss. Although BERT also shows significant improvements, it is still less effective than RoBERTa and DeBERTa. This difference in performance may be attributed to the more complex attention mechanisms and extended training steps utilized by DeBERTa and RoBERTa, enhancing their capabilities.

\revision{The distilled models still exhibit lower accuracy and higher loss on the testing sets compared to the training sets. For example, BERT achieves 93.6\% accuracy and 0.03 loss on the training set, while its accuracy slightly drops to 91.90\%, and the loss increases to 0.05 on the test set. Similarly, RoBERTa's accuracy decreases from 96.73\% on the training set to 93.60\% on the test set, and its loss rises from 0.01 to 0.02. While this discrepancy in performance between the two sets suggests potential overfitting to the training data, the overall performance remains acceptable for these models.}

\begin{tcolorbox}[title=Answer to RQ1 ,boxrule=1pt,boxsep=1pt,left=2pt,right=2pt,top=2pt,bottom=2pt]
\tool{} effectively utilizes knowledge distillation to learn from LLMs, proficiently classifying queries that are likely to produce toxic content.
\end{tcolorbox}

\begin{figure}[!htpb]
    \centering
    \includegraphics[width=0.35\textwidth]{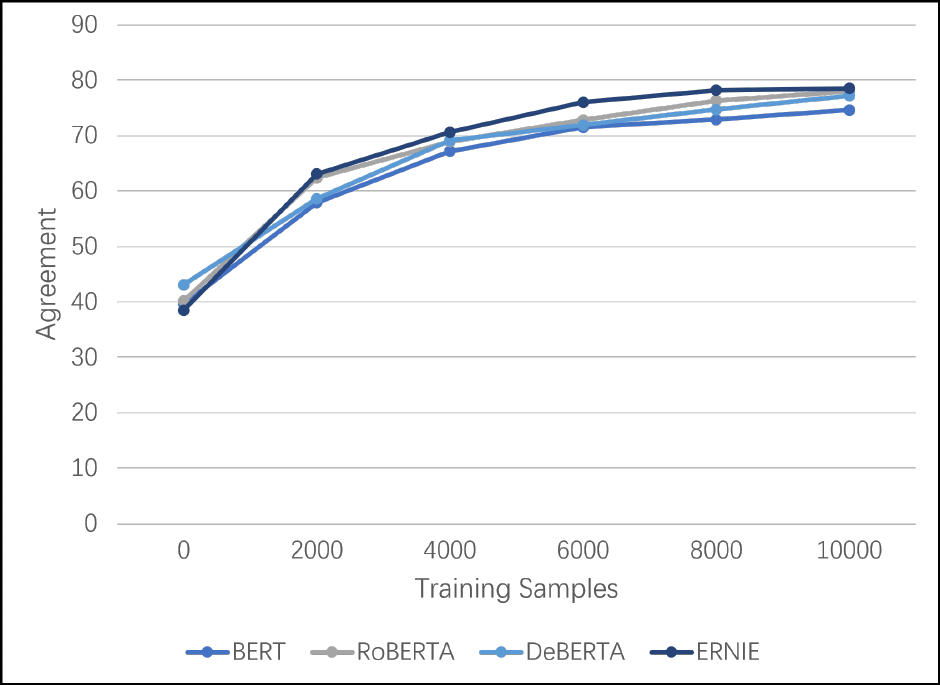}
    \caption{Distilled Model Agreement with GPT-3.5 Curve with Varying Training Sample Sizes.}
    \label{fig:agreement}
\end{figure}

\subsection{RQ2: Tradeoff}
\noindent\textbf{Motivation.} 
In this research question, we study the trade-off between model performance and dataset size, as our goal is to optimize the cost of testing LLMs. Although scaling up the dataset can improve model performance, it also leads to higher training costs. In this research question, we focus on determining the ideal dataset size for cost-effective testing.

\noindent\textbf{Method.}  \revision{To investigate the impact of training dataset size on the performance of the distilled model, we conduct a series of experiments with varying sample sizes. Starting from 0 samples, we incrementally increase the number of samples by 2,000 in each iteration. At each step, we train the distilled model using the corresponding number of samples and evaluate its agreement with GPT-3.5 on the 1,000 randomly selected toxic comments from the RealToxicPrompts dataset. To represent a real testing scenario, we label the queries based directly on the responses of GPT-3.5. Since these responses can be influenced by the variability of the LLM, the agreement result will decrease to some extent. }

\noindent\textbf{Results.} 
\revision{ 
As depicted in Figure~\ref{fig:agreement}, the performance of our framework improves with an increase in sample size. However, the growth rate of agreement decreases as the sample size increases. }

\revision{We assume that one query requires 200 tokens, and the LLM's output uses 100 tokens. According to OpenAI's pricing scheme~\cite{gptpricing}, it can be determined that for every 2,000 queries, the cost would be approximately \$8. Specifically, the average agreement improvement per 2,000 additional samples decreases from 28.28\% (from 0 to 2,000 samples) to a mere 1.55\% (from 8,000 to 10,000 samples). The average cost per 1 percentage point improvement in agreement escalates from \$0.28 (from 0 to 2,000 samples) to a staggering \$10.32 (from 8,000 to 10,000 samples).}

\revision{ When evaluating the practicality of a model, it is crucial to consider both the training costs and the costs associated with tests. For instance, training a model using 4,000 samples incurred a data collection cost of \$32 and a training cost of \$0.35, resulting in a total cost of \$32.35. The model achieves an average agreement of around 70\%. It costs \$0.005 to identify an effectively malicious query. Assuming we need to identify 10,000 effectively malicious queries, the total cost would be \$82.35. However, training the model with 2,000 samples and 6,000 samples would incur costs of \$91 and \$92.8, respectively. In this section, considering the slow growth in agreement scores relative to the increase in training samples, we will evaluate our approach using 4,000 samples during the knowledge distillation phase. In a real testing scenario, testers may select the appropriate data size for knowledge distillation based on the size of their testing dataset. }

\begin{tcolorbox}[title=Answer to RQ2 ,boxrule=1pt,boxsep=1pt,left=2pt,right=2pt,top=2pt,bottom=2pt]
While the effectiveness of the model improves with an increased amount of data used during knowledge distillation, the growth rate slows down later on. Taking cost into account, we choose to use 4,000 samples for knowledge distillation.
\end{tcolorbox}

\begin{table*}[!htpb]
\centering
\caption{Agreement of Models Before Knowledge Distillation (B.K.D.) and After Knowledge Distillation (A.K.D.) with LLMs}
\label{tab:agreement}
\scalebox{0.98}{

\begin{tabular}{l|cc|cc|cc|cc|cc}
\toprule
\multirow{2}{*}{\diagbox{Model}{LLM}} & \multicolumn{2}{c|}{GPT-3.5} & \multicolumn{2}{c|}{GPT-4.0} & \multicolumn{2}{c|}{Vicuna-13B} & \multicolumn{2}{c|}{LLama-13B} & \multicolumn{2}{c}{Average Agreement} \\
                                      & B.K.D     & A.K.D.              & B.K.D     & A.K.D.              & B.K.D       & A.K.D.               & B.K.D      & A.K.D.                            & B.K.D          & A.K.D.                  \\ \hline
BERT    &    17.30               & 66.50          &    13.40       & 57.60          & \textbf{25.60}  & 61.90               &   16.80        & 63.50               & 18.28          & 62.38                   \\
RoBERTa &    \textbf{33.60}      & 69.70          & \textbf{24.20} & 63.00          &    14.40        & \textbf{68.50}      &   29.50        & 62.30               & 25.43          & 65.88                   \\
DeBERTa &    26.70               & 68.50          &    19.60       & \textbf{64.80} &    19.50        & 65.10               &   20.10        & \textbf{68.20}      & 21.48          & \textbf{66.65}          \\
ERNIE   &    24.50               & \textbf{71.80} &    21.50       & 59.30          &    23.20        & 55.10               & \textbf{34.60} & 52.10               & \textbf{25.95} & 59.58                   \\ \bottomrule
\end{tabular}
}
\end{table*}

\begin{table}[!htpb]
    \caption{The Loss of the Distilled Model in Jigsaw Dataset}
    \label{tab:jigsaw_loss}
    \scalebox{0.98}{
    \begin{tabular}{lcc}
    \toprule[1.5pt]
    \multirow{2}{*}{Model} & \multicolumn{2}{c}{Evaluation Result}                       \\ \cline{2-3} 
                           & B.K.D. Loss & A.K.D. Loss \\ \hline
    BERT                   & 0.78        & 0.17        \\
    RoBERTa                & 0.76        & 0.12        \\
    DeBERTa                & 0.64        & 0.19        \\
    ERNIE                  & 0.85        & 0.23        \\ \bottomrule[1.5pt]
    \end{tabular}
}
    
\end{table}

\subsection{RQ3: Generality}
\noindent\textbf{Motivation.} 
In this research question, our goal is to assess whether our approach performs effectively when applied to various types of LLMs and when utilizing new data from a different dataset. The reason for this evaluation is that given the rapid development of large models, a method must be effective across different large models. Furthermore, while our approach performs well on our dataset, it is meaningful to test it in different scenarios since the data within the same dataset may exhibit some similarities.

\noindent\textbf{Method.}

In this research question, we evaluate our framework's performance on four popular LLMs: GPT-3.5~\cite{gpt3.5}, GPT-4~\cite{gpt4}, Vicuna-13B~\cite{vicuna}, and LLama-13B~\cite{llama}. To mirror real-world scenarios, we use the default settings of these LLMs, including temperature and top-p values, for response generation. During the training phase, we select 4,000 samples for knowledge distillation. We employ the same method as in RQ1 to train the classification models.

For the new data, we utilize a new dataset from Kaggle's Jigsaw Toxic Comment Classification Challenge\footnote{\url{https://kaggle.com/competitions/jigsaw-multilingual-toxic-comment-classification}}. This dataset comprises Wikipedia comments tagged for toxicity by human reviewers. We randomly selected 1,000 queries from the Jigsaw dataset for evaluation. We use the same method in RQ1 to calculate the agreement.

\noindent\textbf{Results.} The evaluation results are detailed in Table \ref{tab:agreement}. The results show that the agreement between the distilled models and the LLMs under test ranges from 52.10\% to 71.80\%. DeBERTa exhibits the highest average agreement with the LLMs under test at 66.65\%, while ERNIE shows the lowest agreement at 59.58\%. After knowledge distillation, all distilled models show an improvement of approximately 40\% in their agreement with the classifications of the LLMs under test. This underlines the effectiveness of knowledge distillation in enhancing the distilled models' ability to recognize and address malicious queries.
It is noteworthy that different LLMs exhibit varying degrees of alignment with the best-performing distilled model. For instance, Vicuna-13B aligns best with RoBERTa, while the LLama series shows more agreement with BERT and DeBERTa. These differences likely stem from the architectural similarities between the moderation components of the LLMs and the distilled models; a closer architectural resemblance leads to higher agreement. The vulnerable moderation mechanism of GPT-3.5 enables the distilled models to attain a higher average agreement of 69.12\%.

Some distilled models exhibit notably high agreement with the LLMs under test. For example, the agreement peaks at 71.80\% between ERNIE and GPT-3.5, and 64.80\% between DeBERTa and GPT-4.0. These findings indicate that newer LLMs, such as GPT-4.0, exhibit less agreement with the distilled models compared to earlier versions like GPT-3.5, reflecting the advancements in the safety mechanisms of LLMs.

Compared to the results of Table~\ref{tab:combine_agreement}, the distilled model shows a decrease in agreement with the data from the Jigsaw dataset. We believe this decrease can be attributed to several factors. First, the distilled model might have undergone overfitting during its training process, leading to subpar performance when encountering new data. Second, the inherent randomness of LLMs means that they may provide different responses to the same malicious content upon multiple queries. This results in the distilled model's inability to make accurate predictions. Third, since we employed response-based knowledge distillation, there was insufficient utilization of the internal information of the LLM, leading to poor performance in the agreement metric. Despite this, the average enhancement in agreement after knowledge distillation is 63.74\%, with individual models showing distinct improvement patterns. 

\revision{To further investigate the performance degradation, we present the loss values obtained from these models on the Jigsaw dataset in Table ~\ref{tab:jigsaw_loss}.  Upon comparing the results in Table ~\ref{tab:combine_agreement} and \ref{tab:jigsaw_loss}, we observe that while the knowledge distillation process reduces the loss for the Jigsaw dataset, the loss remains higher when compared to the loss data in Table ~\ref{tab:combine_agreement}. This suggests that the model may rely on specific characteristics unique to the training data. This also highlights that our method can be enhanced by incorporating additional datasets or augmenting the existing dataset with more diverse examples. }

\begin{tcolorbox}[title=Answer to RQ3,boxrule=1pt,boxsep=1pt,left=2pt,right=2pt,top=2pt,bottom=2pt]

While the inherent randomness of large models can influence our method's performance, the results indicate that our approach can effectively learn the moderation mechanisms of various LLMs across different contexts.

\end{tcolorbox}

\subsection{RQ4: Comparisons}

\noindent\textbf{Motivation.} In this research question, we evaluate the effectiveness of \tool{} in testing LLMs. We compare the Attack Success Rates (ASR) of our approach with state-of-the-art tools, showing that our approach can lead to cost savings in testing. In addition, we also integrate state-of-the-art tools with our filters, illustrating how our approach can seamlessly complement well-established tools.

\noindent\textbf{Method.} To evaluate our approach, we compare its performance with two baseline methods, random word replacement and Textfooler, and state-of-the-art tools including JailbreakingLLMs~\cite{chao2023jailbreaking}, GPTFuzzer~\cite{yu2023gptfuzzer}, FuzzLLM~\cite{yao2023fuzzllm}, HouYi~\cite{liu2023prompt}. Specifically, we utilize these methods to generate 1,000 queries that may potentially elicit a toxic response and feed them to the LLMs under testing. We assess their efficacy by calculating the ASR, which is calculated as:
$$ASR = \frac{\text{Number of effective queries}}{\text{Total number of queries}}$$

To enhance the evaluation of our knowledge distillation approach, we employ the DeBERTa models, selected for their outstanding performance according to the results of RQ3, as filters.  For each approach, we filter out ineffective queries before presenting them to the LLMs for testing. In this case,  ASR is calculated as:
$$ASR = \frac{\text{Number of effective queries}}{\text{Number of queries passing the filter}}$$

\begin{table*}
    \centering
    \caption{Attack Success Rates (ASRs) of different malicious query generation methods on LLMs}
    \label{tab:asr}

    \begin{tabular}{ccccccc}
    \toprule[1.5pt]
\multicolumn{2}{c}{\multirow{2}{*}{Method}}        & \multicolumn{5}{c}{LLMs}                           \\ \cline{3-7} 
\multicolumn{2}{c}{}                               & GPT 3.5 & GPT 4 & Vicuna-13B & LLama-13B & Average \\ \hline
\multirow{2}{*}{Random}            & Without Filter & 2.20     & 1.70   & 1.20        & 0.80       & 1.48   \\
                                   & Filter         & 44.30    & 37.20  & 40.40       & 39.50      & 40.35   \\ \hline
\multirow{2}{*}{Textfooler}        & Without Filter & 15.30    & 10.70  & 14.30       & 17.40      & 14.43  \\
                                   & Filter         & 36.90    & 42.10  & 40.30       & 33.30      & 38.15   \\ \hline
\multirow{2}{*}{\textbf{Syntax-tree-based}} & Without Filter & 23.60    & 18.90  & 24.10       & 28.30      & 23.73  \\
                                   & Filter         & 56.30    & 48.50  & 45.50       & 48.80      & 49.78  \\ \hline
\multirow{2}{*}{\textbf{LLM-based}}         & Without Filter & 39.30    & 23.80  & 32.40       & 32.50      & 32.00      \\
                                   & Filter         & \textbf{60.70}   & \textbf{52.90}  & \textbf{59.50}      & \textbf{59.90}      & \textbf{58.25}   \\ \hline
\multirow{2}{*}{JailbreakingLLMs}  & Without Filter & 15.90    & 12.30  & 10.60      & 14.50     & 13.33  \\
                                   & Filter         & 44.00   & 36.00 & 34.80      & 41.30     & 39.03  \\ \hline
\multirow{2}{*}{GPTFuzzer}         & Without Filter & 15.40    & 12.30  & 27.30       & 16.50     & 17.88 \\
                                   & Filter         & 36.70   & 30.20 & 48.80      & 31.80     & 36.88 \\ \hline
\multirow{2}{*}{FuzzLLM}           & Without Filter & 8.80     & 5.90   & 19.90       & 16.60      & 12.80    \\
                                   & Filter         & 18.50   & 16.30 & 52.10      & 38.90     & 31.45   \\ \hline
\multirow{2}{*}{MasterKey}         & Without Filter & 18.40    & 11.20  & 19.30       & 15.00        & 15.98  \\
                                   & Filter         & 45.20    & 47.30  & 32.60      & 36.10     & 40.30 \\ \hline
\multirow{2}{*}{HouYi}             & Without Filter & 5.30     & 1.90   & 18.60       & 11.50      & 9.33   \\
                                   & Filter         & 13.90   & 5.30  & 47.30      & 27.60      & 23.53   \\ \bottomrule[1.5pt]
    \end{tabular}

\end{table*}

\noindent\textbf{Results.} The evaluation results are detailed in Table \ref{tab:asr}. The table delineates the performance of \tool{} across two modes: syntax-based and LLM-based methods with a filter. Compared with other methods, the results demonstrate that \tool{} using the LLM-based method with a filter achieves the highest ASR for all types of LLM. Specifically, the LLM-based method with a filter obtains the highest average ASR at 58.25\%, closely followed by the syntax-based method with a filter at 49.78\%. In contrast, other methods without a filter cannot even achieve a 20\% average ASR. This highlights that existing methods focus on generating sophisticated malicious queries without adequately considering the associated costs of interacting with the LLM.

To comprehensively evaluate the influence of filters on the ASRs of each method, we conducted a comparison between results obtained with and without filters in place. The result reveals a significant increase in ASRs across all methods when the corresponding filter is applied. For instance, the Random method experiences a dramatic rise in ASR from 1.48\% without filtering to 40.35\% with filtering, while Textfooler jumps from 14.43\% to 38.15\%. The other methods also exhibit varying degrees of improvement when filtering is employed.  These findings highlight the effectiveness of our knowledge distillation process in creating a distilled model capable of accurately identifying and filtering out invalid or irrelevant malicious queries. Furthermore, this demonstrates that our method can be easily applied to complement existing methods, thereby reducing costs.

Finally, we also compare the performance of different methods against various LLMs.  The results indicate that GPT-4 generally demonstrates the strongest resistance. Conversely, Vicuna-13B and LLama-13B appear to be more susceptible to attacks, particularly when confronted with methods like FuzzLLM and HouYi. However, it is essential to note that this trend is not entirely consistent across all attack methods. For instance, when faced with the LLM-based method, GPT-3.5 exhibits a higher ASR (60.70\%) compared to Vicuna-13B (59.5\%) and LLama-13B (59.9\%). This suggests that the effectiveness of different attack methods can vary depending on the specific LLM being targeted, highlighting the complexity of the interplay between attack methods and LLM architectures. Furthermore, even the relatively robust GPT-4 yields concerningly high ASRs of 48.50\% and 52.90\% under filtered syntax-tree-based and LLM-based attacks, respectively. This highlights the importance of further research and development efforts aimed at enhancing the defense capabilities of LLMs against increasingly effective malicious queries.

\begin{table*}[!htpb]
\caption{ASR for for different top-p}
\label{tab:top-p}
\scalebox{0.95}{
\begin{tabular}{ccccccccc}
\toprule[1.5pt]
\multirow{2}{*}{Top-p} & \multicolumn{2}{c}{GPT 3.5} & \multicolumn{2}{c}{GPT 4} & \multicolumn{2}{c}{Vicuna-13B} & \multicolumn{2}{c}{LLama-13B} \\ \cline{2-9}
                       & Syntax Tree      & LLM      & Syntax Tree     & LLM     & Syntax Tree        & LLM       & Syntax Tree       & LLM       \\ \hline
0.25 & 44.80 & 58.20 & 29.70 & \textbf{53.40} & 36.90 & 57.10 & 43.50 & 58.60 \\
0.5 & 46.20 & 59.50 & 35.80 & 51.70 & 40.50 & 58.30 & \textbf{58.90} & \textbf{ 60.20} \\
0.75 & \textbf{55.30} & \textbf{61.10} & 41.30 & 52.30 & \textbf{41.80} & \textbf{60.80} & 44.00 & 59.10 \\
1 & 51.90 & 60.70 & \textbf{42.40} & 52.90 & 39.60 & 59.50 & 49.40 & 59.90 \\ \bottomrule[1.5pt]
\end{tabular}
}
\end{table*}

\begin{table}[!htpb]
\caption{ASR(\%) for different temperatures on GPTs}
\label{tab:temperatureGPT}
\scalebox{0.95}{
\begin{tabular}{ccccc}
\toprule[1.5pt]
\multicolumn{1}{l}{\multirow{2}{*}{Temperature}} & \multicolumn{2}{c}{GPT 3.5} & \multicolumn{2}{c}{GPT 4} \\ \cline{2-5} 
\multicolumn{1}{l}{}                             & Syntax Tree      & LLM      & Syntax Tree     & LLM     \\ \hline 
0                                                & 21.80             &   15.60      & 10.10          &   5.40  \\
0.5                                              & 40.30             &   45.20      & 36.50          &   39.20   \\
1                                                & 56.30             &   60.70      & 42.40          &   52.90   \\
1.5                                              & 54.30             &\textbf{67.60}& \textbf{53.70} &   53.20   \\
2                                                & \textbf{59.80}    &   62.02      & 48.90          & \textbf{56.10}  \\ \bottomrule[1.5pt]
\end{tabular}
}
\end{table}

\begin{table}[!htpb]
\caption{ASR(\%) for different temperatures on Vicuna and LLama}
\label{tab:temperatureOther}
\scalebox{0.95}{
\begin{tabular}{ccccc}
\toprule[1.5pt]
\multicolumn{1}{l}{\multirow{2}{*}{Temperature}} & \multicolumn{2}{c}{Vicuna-13B} & \multicolumn{2}{c}{LLama-13B} \\ \cline{2-5} 
\multicolumn{1}{l}{}                             & Syntax Tree      & LLM      & Syntax Tree     & LLM     \\ \hline 
0.01                                             & 12.50         &   18.20       & 9.30             &   14.70      \\
0.25                                             & 28.00           &   35.60       & 35.50            &   41.20      \\
0.5                                              & \textbf{45.50}&   59.50       & 52.30            &   59.90      \\
0.75                                             & 39.60         &   55.10       & 57.50            &   65.40      \\
1                                                & 42.10         &\textbf{60.90} & \textbf{60.10}   &\textbf{66.30} \\ \bottomrule[1.5pt]
\end{tabular}
}
\end{table}

\begin{tcolorbox}[title=Answer to RQ4 ,boxrule=1pt,boxsep=1pt,left=2pt,right=2pt,top=2pt,bottom=2pt]
\tool{} demonstrates good performance in testing various LLMs compared to existing methods. The knowledge distillation in our approach can seamlessly complement well-established tools, thereby reducing the cost of testing the LLMs.
\end{tcolorbox}

\subsection{RQ5: Configuration }

\noindent\textbf{Motivation.}
Huang et al.~\cite{huang2023catastrophic} pointed out that altering decoding hyperparameters or sampling methods can significantly impact the reliability of alignment procedures and evaluations in LLMs. In this research question, we explore how different parameters of the LLM impact the effectiveness of our approach.

\noindent\textbf{Method.}
We adjust the top-p values and temperature parameters across four LLMs under evaluation. Specifically, we set temperature ranges from 0 to 2 for GPT models and 0 to 1 for LLama and Vicuna-13B, while the top-p values range from 0 to 1 for all models. Each model undergoes testing under various settings to assess the impact of these parameters on the effectiveness of both Syntax tree-based and LLM-based methods. To facilitate this evaluation, we employ the Syntax tree-based and LLM-based methods to generate 1,000 queries for each parameter setting.

\noindent\textbf{Results.} 
The ASR for different top-p values on LLMs is presented in Table~\ref{tab:top-p}. The results reveal that the effect of top-p on ASR differs among the LLMs tested, with each model exhibiting unique patterns of vulnerability. GPT 3.5 and Vicuna-13B show similar trends, with the highest ASR observed at a top-p value of 0.75 for both the Syntax Tree and LLM-based methods. For GPT 3.5, the ASR reaches 55.30\% and 61.10\% for the Syntax Tree and LLM-based methods, respectively, while Vicuna-13B achieves an ASR of 41.80\% and 60.80\% for the same methods. These results suggest that a more diverse pool of candidate tokens leads to increased vulnerability to attacks in these models. In contrast, GPT 4 demonstrates a less pronounced impact of top-p on ASR. The highest ASR for GPT 4 is achieved at a top-p value of 0.25 for the LLM-based method at 53.40\% and at a top-p value of 1 for the Syntax Tree method at 42.40\%. LLama-13B exhibits a unique pattern, with the highest ASR achieved at a top-p value of 0.5 for both the Syntax Tree and LLM-based methods, reaching 58.90\% and 60.20\%, respectively. This suggests that a moderate level of diversity in the candidate token pool leads to increased vulnerability in LLama-13B, in contrast to the other models tested.

Tables~\ref{tab:temperatureGPT} and~\ref{tab:temperatureOther} present the ASR for different temperature settings on LLMs, evaluated using both Syntax Tree and LLM-based methods. The results indicate that the temperature parameter affects the ASR for all tested LLMs, generally resulting in increased vulnerability to attacks with higher temperatures. For GPT 3.5 and GPT 4, the highest ASR is observed at different temperature settings depending on the evaluation method. For GPT 3.5, the highest ASR is observed at a temperature of 2 for the Syntax Tree method at 59.80\% and a temperature of 1.5 for the LLM-based method at 67.60\%. Similarly, GPT 4 reaches its peak ASR at a temperature of 1.5 for the Syntax Tree method at 53.70\% and at a temperature of 2 for the LLM-based method at 56.10\%. These findings suggest that higher temperature settings make GPT models more vulnerable to attacks compared to lower temperature settings. Vicuna-13B and LLama-13B show a similar trend to GPT models. Vicuna-13B achieves the highest ASR at a temperature of 0.5 for the Syntax Tree method at 45.50\% and a temperature of 1 for the LLM-based method at 60.90\%. LLama-13B reaches its peak ASR at a temperature of 1 for both the Syntax Tree method at 60.10\% and the LLM-based method at 66.30\%. These results indicate that higher temperature settings also make Vicuna-13B and LLama-13B more vulnerable to attacks, consistent with the findings for GPT models.

\begin{tcolorbox}[title=Answer to RQ5 ,boxrule=1pt,boxsep=1pt,left=2pt,right=2pt,top=2pt,bottom=2pt]
The performance of our approach is influenced by hyperparameters such as temperature and top-p values. As the impact of top-p values is generally less significant compared to temperature, we can enhance the ASR by configuring the LLMs with higher temperatures.
\end{tcolorbox}

\subsection{Threats to Validity}

\noindent\textbf{Internal Validity.}
In our evaluation, the internal validity of our findings may be affected by the presence of randomness. A notable challenge arises from the inconsistency observed in LLMs' responses to identical harmful content. To address this inherent unpredictability, we conducted multiple queries for each harmful input as a countermeasure. Furthermore, in the context of knowledge distillation, we limited our selection to four models during our process, recognizing that alternative models may outperform them. Our research primarily investigates the effectiveness of knowledge distillation in enhancing testing efficiency rather than determining optimal model selections. Consequently, our approach may require additional validation and adaptation when applied to different LLMs or other prospective models considered in future research.

\noindent\textbf{External Validity.}
Two primary challenges come to the forefront when considering the aspect of external validity. Firstly, the question arises whether our approach can be effectively extended to the realm of testing with alternative LLMs. Our investigation encompassed a spectrum of widely used LLMs. Variations exist among them in terms of their architecture and training methodologies. While our methodology takes these variations into account, it is imperative to acknowledge that specific adaptations may be necessitated to cater to the unique attributes of each LLM, thus ensuring the preservation of the approach's efficacy. Secondly, in the context of knowledge distillation, we employed the Realtoxicityprompts dataset. The utilization of this particular dataset may impose limitations on the model's ability to handle a diverse range of malicious query inputs. To address this limitation, we introduce the Jigsaw dataset in RQ2. This strategic integration is aimed at enhancing the model's adaptability across a broader spectrum of data types, thus fortifying its capabilities in processing various forms of malicious content. The selection of these supplementary datasets was guided by the objective of encompassing a wider array of thematic and stylistic nuances, thereby bolstering the model's generalization capabilities across an array of distinct scenarios.

\section{Related Work}\label{sec:relatedwork}
\noindent\textbf{LLMs alignment.} 

\revision{Hendrycks et al.~\cite{hendrycks2020aligning} introduced the ETHICS dataset to evaluate LLMs' understanding of ethics, revealing that LLMs have a partial understanding of human ethics but lack comprehensive ethical reasoning capabilities. While this dataset provides a valuable benchmark, it does not address the challenges of efficiently testing LLMs for safety alignment. Dai et al.~\cite{safe-rlhf} developed Beaver, an open-source RLHF framework that provides training data and reproducible code for alignment studies. Although Beaver facilitates the training process, it does not focus on optimizing the testing phase, which is the primary concern of our proposed framework. Sun et al.~\cite{sun2023safety} introduced ``SELF-ALIGN,'' a method enabling the self-adjustment of AI agents with minimal human supervision. While this method ensures that LLM outputs align with human intentions, it does not explicitly address the issue of efficient testing for safety alignment.}

\revision{In contrast to previous works, we introduce a novel framework that aims at the efficient testing of LLMs for safety alignment. By transferring LLMs' moderation knowledge to a small model and employing syntax tree parsing and LLM-based methods for generating malicious queries, we significantly reduce the computational resources required for thorough LLM evaluation.}

\noindent\textbf{Attack on LLMs.} \revision{Aligning LLMs with human values remains a challenging task, as demonstrated by recent research~\cite{liu2023autodan, shen2023anything, cao2023defending,deng2024pandorajailbreakgptsretrieval,liu2024grootadversarialtestinggenerative}. Various studies have explored the vulnerabilities of LLMs and proposed methods to exploit or defend against these weaknesses~\cite{huang2024semantic,li2024halluvaul, li2024lock}. Wallace et al.~\cite{wallace-etal-2019-universal} showed that contextual triggers can lead to biased and harmful LLM responses, highlighting the need for robust safety measures. Gupta et al.~\cite{gupta2023chatgpt} investigated methods like jailbreaking and prompt injection to bypass LLM ethical constraints, demonstrating the potential for adversarial actors to exploit LLMs for malicious activities. However, their work focused on manual exploitation techniques, whereas our approach automates the testing process to identify vulnerabilities more efficiently. SelfCipher~\cite{yuan2023gpt} uses natural language role-playing to unlock LLMs' hidden encryption capabilities, circumventing safety protocols. While their work showcases the potential for encrypted messaging to evade LLM safety measures, our approach tackles the issue from a different angle by generating targeted test cases to identify and address vulnerabilities. Perez et al.~\cite{perez2022ignore} proposed PromptInject, demonstrating that simple handcrafted inputs can easily misalign GPT-3. Their work highlights the need for comprehensive testing of LLMs. In contrast, our approach leverages an automatic framework to generate a diverse set of test cases, reducing the reliance on manual input crafting and increasing the efficiency of the testing process. BadPrompt~\cite{cai2022badprompt} is a lightweight and task-adaptive algorithm for backdoor attacks on continuous prompts, effectively attacking while maintaining high performance on clean test sets. PPT~\cite{du2022ppt} implants malicious behaviors into prompts, achieving a 99\% attack success rate on pre-trained language models with minimal accuracy sacrifice on the original task. Unlike these white-box approaches that rely on access to or manipulation of training data, our method focuses on a more realistic black-box scenario, generating effective attacks only based on the model's output.}

\section{Conclusion } \label{sec:conclusion}
In this paper, we introduce a novel framework named \tool{}, designed to minimize the cost of testing LLMs. Our approach combines knowledge distillation and prompt engineering to automatically generate test cases that provoke LLMs to output toxic content. Additionally, through experiments, we demonstrate the effectiveness of our approach across several mainstream LLMs, surpassing the capabilities of state-of-the-art tools in terms of testing cost efficiency. 
We also emphasize that our approach can effectively complement various attack strategies. Testers can enhance their efficiency when testing LLMs by simply utilizing our approach to train a filter. In future work, we plan to explore whether access to neural information can lead to the development of a filter with superior performance and efficiency.

\section*{Acknowledgements}
We sincerely thank all the anonymous reviewers for their valuable feedback, which greatly contributed to the improvement of this paper. This research was jointly sponsored by the NSFC Program under Grants No. 62302304 and the ShanghaiTech Startup Funding.

\bibliographystyle{ACM-Reference-Format}
\balance
\bibliography{reference}


\begin{thebibliography}{45}

%%% the Web address.

\ifx \showCODEN    \undefined \def \showCODEN     #1{\unskip}     \fi
\ifx \showDOI      \undefined \def \showDOI       #1{#1}\fi
\ifx \showISBNx    \undefined \def \showISBNx     #1{\unskip}     \fi
\ifx \showISBNxiii \undefined \def \showISBNxiii  #1{\unskip}     \fi
\ifx \showISSN     \undefined \def \showISSN      #1{\unskip}     \fi
\ifx \showLCCN     \undefined \def \showLCCN      #1{\unskip}     \fi
\ifx \shownote     \undefined \def \shownote      #1{#1}          \fi
\ifx \showarticletitle \undefined \def \showarticletitle #1{#1}   \fi
\ifx \showURL      \undefined \def \showURL       {\relax}        \fi
% invisible to TeX
\providecommand\bibfield[2]{#2}
\providecommand\bibinfo[2]{#2}
\providecommand\natexlab[1]{#1}
\providecommand\showeprint[2][]{arXiv:#2}

\bibitem[andyll7772(2023)]%
        {vicuna13}
\bibfield{author}{\bibinfo{person}{andyll7772}.} \bibinfo{year}{2023}\natexlab{}.
\newblock \bibinfo{title}{Run a Chatgpt-like Chatbot on a Single GPU with ROCm}.
\newblock \bibinfo{howpublished}{\url{https://github.com/huggingface/blog/blob/main/chatbot-amd-gpu.md}}.
\newblock


\bibitem[Cai et~al\mbox{.}(2022)]%
        {cai2022badprompt}
\bibfield{author}{\bibinfo{person}{Xiangrui Cai}, \bibinfo{person}{Haidong Xu}, \bibinfo{person}{Sihan Xu}, \bibinfo{person}{Ying Zhang}, {et~al\mbox{.}}} \bibinfo{year}{2022}\natexlab{}.
\newblock \showarticletitle{Badprompt: Backdoor attacks on continuous prompts}.
\newblock \bibinfo{journal}{\emph{Advances in Neural Information Processing Systems}}  \bibinfo{volume}{35} (\bibinfo{year}{2022}), \bibinfo{pages}{37068--37080}.
\newblock


\bibitem[Cao et~al\mbox{.}(2023)]%
        {cao2023defending}
\bibfield{author}{\bibinfo{person}{Bochuan Cao}, \bibinfo{person}{Yuanpu Cao}, \bibinfo{person}{Lu Lin}, {and} \bibinfo{person}{Jinghui Chen}.} \bibinfo{year}{2023}\natexlab{}.
\newblock \showarticletitle{Defending against alignment-breaking attacks via robustly aligned llm}.
\newblock \bibinfo{journal}{\emph{arXiv preprint arXiv:2309.14348}} (\bibinfo{year}{2023}).
\newblock


\bibitem[Chang et~al\mbox{.}(2024)]%
        {chang2024playguessinggamellm}
\bibfield{author}{\bibinfo{person}{Zhiyuan Chang}, \bibinfo{person}{Mingyang Li}, \bibinfo{person}{Yi Liu}, \bibinfo{person}{Junjie Wang}, \bibinfo{person}{Qing Wang}, {and} \bibinfo{person}{Yang Liu}.} \bibinfo{year}{2024}\natexlab{}.
\newblock \bibinfo{title}{Play Guessing Game with LLM: Indirect Jailbreak Attack with Implicit Clues}.
\newblock
\newblock
\showeprint[arxiv]{2402.09091}~[cs.CR]
\urldef\tempurl%
\url{https://arxiv.org/abs/2402.09091}
\showURL{%
\tempurl}


\bibitem[Chao et~al\mbox{.}(2023)]%
        {chao2023jailbreaking}
\bibfield{author}{\bibinfo{person}{Patrick Chao}, \bibinfo{person}{Alexander Robey}, \bibinfo{person}{Edgar Dobriban}, \bibinfo{person}{Hamed Hassani}, \bibinfo{person}{George~J Pappas}, {and} \bibinfo{person}{Eric Wong}.} \bibinfo{year}{2023}\natexlab{}.
\newblock \showarticletitle{Jailbreaking black box large language models in twenty queries}.
\newblock \bibinfo{journal}{\emph{arXiv preprint arXiv:2310.08419}} (\bibinfo{year}{2023}).
\newblock


\bibitem[Dai et~al\mbox{.}(2023)]%
        {safe-rlhf}
\bibfield{author}{\bibinfo{person}{Josef Dai}, \bibinfo{person}{Xuehai Pan}, \bibinfo{person}{Ruiyang Sun}, \bibinfo{person}{Jiaming Ji}, \bibinfo{person}{Xinbo Xu}, \bibinfo{person}{Mickel Liu}, \bibinfo{person}{Yizhou Wang}, {and} \bibinfo{person}{Yaodong Yang}.} \bibinfo{year}{2023}\natexlab{}.
\newblock \showarticletitle{Safe RLHF: Safe Reinforcement Learning from Human Feedback}.
\newblock \bibinfo{journal}{\emph{arXiv preprint arXiv:2310.12773}} (\bibinfo{year}{2023}).
\newblock


\bibitem[Deng et~al\mbox{.}(2023)]%
        {deng2023jailbreaker}
\bibfield{author}{\bibinfo{person}{Gelei Deng}, \bibinfo{person}{Yi Liu}, \bibinfo{person}{Yuekang Li}, \bibinfo{person}{Kailong Wang}, \bibinfo{person}{Ying Zhang}, \bibinfo{person}{Zefeng Li}, \bibinfo{person}{Haoyu Wang}, \bibinfo{person}{Tianwei Zhang}, {and} \bibinfo{person}{Yang Liu}.} \bibinfo{year}{2023}\natexlab{}.
\newblock \showarticletitle{Jailbreaker: Automated jailbreak across multiple large language model chatbots}.
\newblock \bibinfo{journal}{\emph{arXiv preprint arXiv:2307.08715}} (\bibinfo{year}{2023}).
\newblock


\bibitem[Deng et~al\mbox{.}(2024)]%
        {deng2024pandorajailbreakgptsretrieval}
\bibfield{author}{\bibinfo{person}{Gelei Deng}, \bibinfo{person}{Yi Liu}, \bibinfo{person}{Kailong Wang}, \bibinfo{person}{Yuekang Li}, \bibinfo{person}{Tianwei Zhang}, {and} \bibinfo{person}{Yang Liu}.} \bibinfo{year}{2024}\natexlab{}.
\newblock \bibinfo{title}{Pandora: Jailbreak GPTs by Retrieval Augmented Generation Poisoning}.
\newblock
\newblock
\showeprint[arxiv]{2402.08416}~[cs.CR]
\urldef\tempurl%
\url{https://arxiv.org/abs/2402.08416}
\showURL{%
\tempurl}


\bibitem[Devlin et~al\mbox{.}(2019)]%
        {devlin-etal-2019-bert}
\bibfield{author}{\bibinfo{person}{Jacob Devlin}, \bibinfo{person}{Ming-Wei Chang}, \bibinfo{person}{Kenton Lee}, {and} \bibinfo{person}{Kristina Toutanova}.} \bibinfo{year}{2019}\natexlab{}.
\newblock \showarticletitle{{BERT}: Pre-training of Deep Bidirectional Transformers for Language Understanding}. In \bibinfo{booktitle}{\emph{Proceedings of the 2019 Conference of the North {A}merican Chapter of the Association for Computational Linguistics: Human Language Technologies, Volume 1 (Long and Short Papers)}}. \bibinfo{publisher}{Association for Computational Linguistics}, \bibinfo{address}{Minneapolis, Minnesota}, \bibinfo{pages}{4171--4186}.
\newblock
\urldef\tempurl%
\url{https://doi.org/10.18653/v1/N19-1423}
\showDOI{\tempurl}


\bibitem[Diao et~al\mbox{.}(2023)]%
        {diao2023lmflow}
\bibfield{author}{\bibinfo{person}{Shizhe Diao}, \bibinfo{person}{Rui Pan}, \bibinfo{person}{Hanze Dong}, \bibinfo{person}{Ka~Shun Shum}, \bibinfo{person}{Jipeng Zhang}, \bibinfo{person}{Wei Xiong}, {and} \bibinfo{person}{Tong Zhang}.} \bibinfo{year}{2023}\natexlab{}.
\newblock \showarticletitle{Lmflow: An extensible toolkit for finetuning and inference of large foundation models}.
\newblock \bibinfo{journal}{\emph{arXiv preprint arXiv:2306.12420}} (\bibinfo{year}{2023}).
\newblock


\bibitem[Du et~al\mbox{.}(2022)]%
        {du2022ppt}
\bibfield{author}{\bibinfo{person}{Wei Du}, \bibinfo{person}{Yichun Zhao}, \bibinfo{person}{Boqun Li}, \bibinfo{person}{Gongshen Liu}, {and} \bibinfo{person}{Shilin Wang}.} \bibinfo{year}{2022}\natexlab{}.
\newblock \showarticletitle{PPT: Backdoor Attacks on Pre-trained Models via Poisoned Prompt Tuning.}. In \bibinfo{booktitle}{\emph{IJCAI}}. \bibinfo{pages}{680--686}.
\newblock


\bibitem[Gehman et~al\mbox{.}(2020)]%
        {gehman2020realtoxicityprompts}
\bibfield{author}{\bibinfo{person}{Samuel Gehman}, \bibinfo{person}{Suchin Gururangan}, \bibinfo{person}{Maarten Sap}, \bibinfo{person}{Yejin Choi}, {and} \bibinfo{person}{Noah~A Smith}.} \bibinfo{year}{2020}\natexlab{}.
\newblock \showarticletitle{Realtoxicityprompts: Evaluating neural toxic degeneration in language models}.
\newblock \bibinfo{journal}{\emph{arXiv preprint arXiv:2009.11462}} (\bibinfo{year}{2020}).
\newblock


\bibitem[Gokaslan and Cohen(2019)]%
        {Gokaslan2019OpenWeb}
\bibfield{author}{\bibinfo{person}{Aaron Gokaslan} {and} \bibinfo{person}{Vanya Cohen}.} \bibinfo{year}{2019}\natexlab{}.
\newblock \bibinfo{title}{OpenWebText Corpus}.
\newblock \bibinfo{howpublished}{\url{http://Skylion007.github.io/OpenWebTextCorpus}}.
\newblock


\bibitem[Gupta et~al\mbox{.}(2023)]%
        {gupta2023chatgpt}
\bibfield{author}{\bibinfo{person}{Maanak Gupta}, \bibinfo{person}{CharanKumar Akiri}, \bibinfo{person}{Kshitiz Aryal}, \bibinfo{person}{Eli Parker}, {and} \bibinfo{person}{Lopamudra Praharaj}.} \bibinfo{year}{2023}\natexlab{}.
\newblock \showarticletitle{From ChatGPT to ThreatGPT: Impact of generative AI in cybersecurity and privacy}.
\newblock \bibinfo{journal}{\emph{IEEE Access}} (\bibinfo{year}{2023}).
\newblock


\bibitem[He et~al\mbox{.}(2020)]%
        {he2020DeBERTa}
\bibfield{author}{\bibinfo{person}{Pengcheng He}, \bibinfo{person}{Xiaodong Liu}, \bibinfo{person}{Jianfeng Gao}, {and} \bibinfo{person}{Weizhu Chen}.} \bibinfo{year}{2020}\natexlab{}.
\newblock \showarticletitle{Deberta: Decoding-enhanced bert with disentangled attention}.
\newblock \bibinfo{journal}{\emph{arXiv preprint arXiv:2006.03654}} (\bibinfo{year}{2020}).
\newblock


\bibitem[Hendrycks et~al\mbox{.}(2020)]%
        {hendrycks2020aligning}
\bibfield{author}{\bibinfo{person}{Dan Hendrycks}, \bibinfo{person}{Collin Burns}, \bibinfo{person}{Steven Basart}, \bibinfo{person}{Andrew Critch}, \bibinfo{person}{Jerry Li}, \bibinfo{person}{Dawn Song}, {and} \bibinfo{person}{Jacob Steinhardt}.} \bibinfo{year}{2020}\natexlab{}.
\newblock \showarticletitle{Aligning ai with shared human values}.
\newblock \bibinfo{journal}{\emph{arXiv preprint arXiv:2008.02275}} (\bibinfo{year}{2020}).
\newblock


\bibitem[Hinton et~al\mbox{.}(2015)]%
        {hinton2015distilling}
\bibfield{author}{\bibinfo{person}{Geoffrey Hinton}, \bibinfo{person}{Oriol Vinyals}, {and} \bibinfo{person}{Jeff Dean}.} \bibinfo{year}{2015}\natexlab{}.
\newblock \showarticletitle{Distilling the knowledge in a neural network}.
\newblock \bibinfo{journal}{\emph{arXiv preprint arXiv:1503.02531}} (\bibinfo{year}{2015}).
\newblock


\bibitem[Huang et~al\mbox{.}(2023)]%
        {huang2023catastrophic}
\bibfield{author}{\bibinfo{person}{Yangsibo Huang}, \bibinfo{person}{Samyak Gupta}, \bibinfo{person}{Mengzhou Xia}, \bibinfo{person}{Kai Li}, {and} \bibinfo{person}{Danqi Chen}.} \bibinfo{year}{2023}\natexlab{}.
\newblock \showarticletitle{Catastrophic jailbreak of open-source LLMs via exploiting generation}.
\newblock \bibinfo{journal}{\emph{arXiv preprint arXiv:2310.06987}} (\bibinfo{year}{2023}).
\newblock


\bibitem[Huang et~al\mbox{.}(2024)]%
        {huang2024semantic}
\bibfield{author}{\bibinfo{person}{Yihao Huang}, \bibinfo{person}{Chong Wang}, \bibinfo{person}{Xiaojun Jia}, \bibinfo{person}{Qing Guo}, \bibinfo{person}{Felix Juefei-Xu}, \bibinfo{person}{Jian Zhang}, \bibinfo{person}{Geguang Pu}, {and} \bibinfo{person}{Yang Liu}.} \bibinfo{year}{2024}\natexlab{}.
\newblock \showarticletitle{Semantic-guided Prompt Organization for Universal Goal Hijacking against LLMs}.
\newblock \bibinfo{journal}{\emph{arXiv preprint arXiv:2405.14189}} (\bibinfo{year}{2024}).
\newblock


\bibitem[Karl and Scherp(2022)]%
        {karl2022transformers}
\bibfield{author}{\bibinfo{person}{Fabian Karl} {and} \bibinfo{person}{Ansgar Scherp}.} \bibinfo{year}{2022}\natexlab{}.
\newblock \showarticletitle{Transformers are Short Text Classifiers: A Study of Inductive Short Text Classifiers on Benchmarks and Real-world Datasets}.
\newblock \bibinfo{journal}{\emph{arXiv preprint arXiv:2211.16878}} (\bibinfo{year}{2022}).
\newblock


\bibitem[Li et~al\mbox{.}(2024c)]%
        {li2024crosslanguageinvestigationjailbreakattacks}
\bibfield{author}{\bibinfo{person}{Jie Li}, \bibinfo{person}{Yi Liu}, \bibinfo{person}{Chongyang Liu}, \bibinfo{person}{Ling Shi}, \bibinfo{person}{Xiaoning Ren}, \bibinfo{person}{Yaowen Zheng}, \bibinfo{person}{Yang Liu}, {and} \bibinfo{person}{Yinxing Xue}.} \bibinfo{year}{2024}\natexlab{c}.
\newblock \bibinfo{title}{A Cross-Language Investigation into Jailbreak Attacks in Large Language Models}.
\newblock
\newblock
\showeprint[arxiv]{2401.16765}~[cs.CR]
\urldef\tempurl%
\url{https://arxiv.org/abs/2401.16765}
\showURL{%
\tempurl}


\bibitem[Li et~al\mbox{.}(2024a)]%
        {li2024halluvaul}
\bibfield{author}{\bibinfo{person}{Ningke Li}, \bibinfo{person}{Yuekang Li}, \bibinfo{person}{Yi Liu}, \bibinfo{person}{Ling Shi}, \bibinfo{person}{Kailong Wang}, {and} \bibinfo{person}{Haoyu Wang}.} \bibinfo{year}{2024}\natexlab{a}.
\newblock \showarticletitle{HalluVault: A Novel Logic Programming-aided Metamorphic Testing Framework for Detecting Fact-Conflicting Hallucinations in Large Language Models}.
\newblock \bibinfo{journal}{\emph{arXiv preprint arXiv:2405.00648}} (\bibinfo{year}{2024}).
\newblock


\bibitem[Li et~al\mbox{.}(2024b)]%
        {li2024lock}
\bibfield{author}{\bibinfo{person}{Yuxi Li}, \bibinfo{person}{Yi Liu}, \bibinfo{person}{Yuekang Li}, \bibinfo{person}{Ling Shi}, \bibinfo{person}{Gelei Deng}, \bibinfo{person}{Shengquan Chen}, {and} \bibinfo{person}{Kailong Wang}.} \bibinfo{year}{2024}\natexlab{b}.
\newblock \showarticletitle{Lockpicking LLMs: A Logit-Based Jailbreak Using Token-level Manipulation}.
\newblock \bibinfo{journal}{\emph{arXiv preprint arXiv:2405.13068}} (\bibinfo{year}{2024}).
\newblock


\bibitem[Liu et~al\mbox{.}(2023c)]%
        {liu2023autodan}
\bibfield{author}{\bibinfo{person}{Xiaogeng Liu}, \bibinfo{person}{Nan Xu}, \bibinfo{person}{Muhao Chen}, {and} \bibinfo{person}{Chaowei Xiao}.} \bibinfo{year}{2023}\natexlab{c}.
\newblock \showarticletitle{Autodan: Generating stealthy jailbreak prompts on aligned large language models}.
\newblock \bibinfo{journal}{\emph{arXiv preprint arXiv:2310.04451}} (\bibinfo{year}{2023}).
\newblock


\bibitem[Liu et~al\mbox{.}(2023a)]%
        {liu2023prompt}
\bibfield{author}{\bibinfo{person}{Yi Liu}, \bibinfo{person}{Gelei Deng}, \bibinfo{person}{Yuekang Li}, \bibinfo{person}{Kailong Wang}, \bibinfo{person}{Tianwei Zhang}, \bibinfo{person}{Yepang Liu}, \bibinfo{person}{Haoyu Wang}, \bibinfo{person}{Yan Zheng}, {and} \bibinfo{person}{Yang Liu}.} \bibinfo{year}{2023}\natexlab{a}.
\newblock \showarticletitle{Prompt Injection attack against LLM-integrated Applications}.
\newblock \bibinfo{journal}{\emph{arXiv preprint arXiv:2306.05499}} (\bibinfo{year}{2023}).
\newblock


\bibitem[Liu et~al\mbox{.}(2023b)]%
        {liu2023jailbreaking}
\bibfield{author}{\bibinfo{person}{Yi Liu}, \bibinfo{person}{Gelei Deng}, \bibinfo{person}{Zhengzi Xu}, \bibinfo{person}{Yuekang Li}, \bibinfo{person}{Yaowen Zheng}, \bibinfo{person}{Ying Zhang}, \bibinfo{person}{Lida Zhao}, \bibinfo{person}{Tianwei Zhang}, {and} \bibinfo{person}{Yang Liu}.} \bibinfo{year}{2023}\natexlab{b}.
\newblock \showarticletitle{Jailbreaking chatgpt via prompt engineering: An empirical study}.
\newblock \bibinfo{journal}{\emph{arXiv preprint arXiv:2305.13860}} (\bibinfo{year}{2023}).
\newblock


\bibitem[Liu et~al\mbox{.}(2019)]%
        {liu2019RoBERTa}
\bibfield{author}{\bibinfo{person}{Yinhan Liu}, \bibinfo{person}{Myle Ott}, \bibinfo{person}{Naman Goyal}, \bibinfo{person}{Jingfei Du}, \bibinfo{person}{Mandar Joshi}, \bibinfo{person}{Danqi Chen}, \bibinfo{person}{Omer Levy}, \bibinfo{person}{Mike Lewis}, \bibinfo{person}{Luke Zettlemoyer}, {and} \bibinfo{person}{Veselin Stoyanov}.} \bibinfo{year}{2019}\natexlab{}.
\newblock \showarticletitle{Roberta: A robustly optimized bert pretraining approach}.
\newblock \bibinfo{journal}{\emph{arXiv preprint arXiv:1907.11692}} (\bibinfo{year}{2019}).
\newblock


\bibitem[Liu et~al\mbox{.}(2024)]%
        {liu2024grootadversarialtestinggenerative}
\bibfield{author}{\bibinfo{person}{Yi Liu}, \bibinfo{person}{Guowei Yang}, \bibinfo{person}{Gelei Deng}, \bibinfo{person}{Feiyue Chen}, \bibinfo{person}{Yuqi Chen}, \bibinfo{person}{Ling Shi}, \bibinfo{person}{Tianwei Zhang}, {and} \bibinfo{person}{Yang Liu}.} \bibinfo{year}{2024}\natexlab{}.
\newblock \bibinfo{title}{Groot: Adversarial Testing for Generative Text-to-Image Models with Tree-based Semantic Transformation}.
\newblock
\newblock
\showeprint[arxiv]{2402.12100}~[cs.CL]
\urldef\tempurl%
\url{https://arxiv.org/abs/2402.12100}
\showURL{%
\tempurl}


\bibitem[Meta(2023)]%
        {llama}
\bibfield{author}{\bibinfo{person}{Meta}.} \bibinfo{year}{2023}\natexlab{}.
\newblock \bibinfo{title}{"LLama-13B"}.
\newblock \bibinfo{howpublished}{\url{https://github.com/facebookresearch/llama/tree/llama_v1}}.
\newblock


\bibitem[OpenAI(2023a)]%
        {gpt3.5}
\bibfield{author}{\bibinfo{person}{OpenAI}.} \bibinfo{year}{2023}\natexlab{a}.
\newblock \bibinfo{title}{"GPT-3.5 Turbo"}.
\newblock \bibinfo{howpublished}{\url{https://platform.openai.com/docs/models/gpt-3-5}}.
\newblock


\bibitem[OpenAI(2023b)]%
        {gpt4}
\bibfield{author}{\bibinfo{person}{OpenAI}.} \bibinfo{year}{2023}\natexlab{b}.
\newblock \bibinfo{title}{"GPT-4"}.
\newblock \bibinfo{howpublished}{\url{https://platform.openai.com/docs/models/gpt-4-and-gpt-4-turbo}}.
\newblock


\bibitem[OpenAI(2023c)]%
        {gptpricing}
\bibfield{author}{\bibinfo{person}{OpenAI}.} \bibinfo{year}{2023}\natexlab{c}.
\newblock \bibinfo{title}{Language models pricing}.
\newblock \bibinfo{howpublished}{\url{https://web.archive.org/web/20231031033745/https://openai.com/pricing}}.
\newblock


\bibitem[Perez and Ribeiro(2022)]%
        {perez2022ignore}
\bibfield{author}{\bibinfo{person}{F{\'a}bio Perez} {and} \bibinfo{person}{Ian Ribeiro}.} \bibinfo{year}{2022}\natexlab{}.
\newblock \showarticletitle{Ignore previous prompt: Attack techniques for language models}.
\newblock \bibinfo{journal}{\emph{arXiv preprint arXiv:2211.09527}} (\bibinfo{year}{2022}).
\newblock


\bibitem[Shen et~al\mbox{.}(2023)]%
        {shen2023anything}
\bibfield{author}{\bibinfo{person}{Xinyue Shen}, \bibinfo{person}{Zeyuan Chen}, \bibinfo{person}{Michael Backes}, \bibinfo{person}{Yun Shen}, {and} \bibinfo{person}{Yang Zhang}.} \bibinfo{year}{2023}\natexlab{}.
\newblock \showarticletitle{" do anything now": Characterizing and evaluating in-the-wild jailbreak prompts on large language models}.
\newblock \bibinfo{journal}{\emph{arXiv preprint arXiv:2308.03825}} (\bibinfo{year}{2023}).
\newblock


\bibitem[Sun et~al\mbox{.}(2023)]%
        {sun2023safety}
\bibfield{author}{\bibinfo{person}{Hao Sun}, \bibinfo{person}{Zhexin Zhang}, \bibinfo{person}{Jiawen Deng}, \bibinfo{person}{Jiale Cheng}, {and} \bibinfo{person}{Minlie Huang}.} \bibinfo{year}{2023}\natexlab{}.
\newblock \showarticletitle{Safety Assessment of Chinese Large Language Models}.
\newblock \bibinfo{journal}{\emph{arXiv preprint arXiv:2304.10436}} (\bibinfo{year}{2023}).
\newblock


\bibitem[Sun et~al\mbox{.}(2020)]%
        {sun2020ernie}
\bibfield{author}{\bibinfo{person}{Yu Sun}, \bibinfo{person}{Shuohuan Wang}, \bibinfo{person}{Yukun Li}, \bibinfo{person}{Shikun Feng}, \bibinfo{person}{Hao Tian}, \bibinfo{person}{Hua Wu}, {and} \bibinfo{person}{Haifeng Wang}.} \bibinfo{year}{2020}\natexlab{}.
\newblock \showarticletitle{Ernie 2.0: A continual pre-training framework for language understanding}. In \bibinfo{booktitle}{\emph{Proceedings of the AAAI conference on artificial intelligence}}, Vol.~\bibinfo{volume}{34}. \bibinfo{pages}{8968--8975}.
\newblock


\bibitem[Team(2023)]%
        {vicuna}
\bibfield{author}{\bibinfo{person}{The~Vicuna Team}.} \bibinfo{year}{2023}\natexlab{}.
\newblock \bibinfo{title}{"Vicuna-13B"}.
\newblock \bibinfo{howpublished}{\url{https://github.com/lm-sys/FastChat}}.
\newblock


\bibitem[Wallace et~al\mbox{.}(2019)]%
        {wallace-etal-2019-universal}
\bibfield{author}{\bibinfo{person}{Eric Wallace}, \bibinfo{person}{Shi Feng}, \bibinfo{person}{Nikhil Kandpal}, \bibinfo{person}{Matt Gardner}, {and} \bibinfo{person}{Sameer Singh}.} \bibinfo{year}{2019}\natexlab{}.
\newblock \showarticletitle{Universal Adversarial Triggers for Attacking and Analyzing {NLP}}. In \bibinfo{booktitle}{\emph{Proceedings of the 2019 Conference on Empirical Methods in Natural Language Processing and the 9th International Joint Conference on Natural Language Processing (EMNLP-IJCNLP)}}. \bibinfo{publisher}{Association for Computational Linguistics}, \bibinfo{address}{Hong Kong, China}, \bibinfo{pages}{2153--2162}.
\newblock
\urldef\tempurl%
\url{https://doi.org/10.18653/v1/D19-1221}
\showDOI{\tempurl}


\bibitem[Wei et~al\mbox{.}(2023)]%
        {wei2023jailbroken}
\bibfield{author}{\bibinfo{person}{Alexander Wei}, \bibinfo{person}{Nika Haghtalab}, {and} \bibinfo{person}{Jacob Steinhardt}.} \bibinfo{year}{2023}\natexlab{}.
\newblock \showarticletitle{Jailbroken: How does llm safety training fail?}
\newblock \bibinfo{journal}{\emph{arXiv preprint arXiv:2307.02483}} (\bibinfo{year}{2023}).
\newblock


\bibitem[Xu et~al\mbox{.}(2024)]%
        {xu2024comprehensivestudyjailbreakattack}
\bibfield{author}{\bibinfo{person}{Zihao Xu}, \bibinfo{person}{Yi Liu}, \bibinfo{person}{Gelei Deng}, \bibinfo{person}{Yuekang Li}, {and} \bibinfo{person}{Stjepan Picek}.} \bibinfo{year}{2024}\natexlab{}.
\newblock \bibinfo{title}{A Comprehensive Study of Jailbreak Attack versus Defense for Large Language Models}.
\newblock
\newblock
\showeprint[arxiv]{2402.13457}~[cs.CR]
\urldef\tempurl%
\url{https://arxiv.org/abs/2402.13457}
\showURL{%
\tempurl}


\bibitem[Yao et~al\mbox{.}(2023)]%
        {yao2023fuzzllm}
\bibfield{author}{\bibinfo{person}{Dongyu Yao}, \bibinfo{person}{Jianshu Zhang}, \bibinfo{person}{Ian~G Harris}, {and} \bibinfo{person}{Marcel Carlsson}.} \bibinfo{year}{2023}\natexlab{}.
\newblock \showarticletitle{Fuzzllm: A novel and universal fuzzing framework for proactively discovering jailbreak vulnerabilities in large language models}.
\newblock \bibinfo{journal}{\emph{arXiv preprint arXiv:2309.05274}} (\bibinfo{year}{2023}).
\newblock


\bibitem[Yu et~al\mbox{.}(2023)]%
        {yu2023gptfuzzer}
\bibfield{author}{\bibinfo{person}{Jiahao Yu}, \bibinfo{person}{Xingwei Lin}, {and} \bibinfo{person}{Xinyu Xing}.} \bibinfo{year}{2023}\natexlab{}.
\newblock \showarticletitle{Gptfuzzer: Red teaming large language models with auto-generated jailbreak prompts}.
\newblock \bibinfo{journal}{\emph{arXiv preprint arXiv:2309.10253}} (\bibinfo{year}{2023}).
\newblock


\bibitem[Yuan et~al\mbox{.}(2023)]%
        {yuan2023gpt}
\bibfield{author}{\bibinfo{person}{Youliang Yuan}, \bibinfo{person}{Wenxiang Jiao}, \bibinfo{person}{Wenxuan Wang}, \bibinfo{person}{Jen-tse Huang}, \bibinfo{person}{Pinjia He}, \bibinfo{person}{Shuming Shi}, {and} \bibinfo{person}{Zhaopeng Tu}.} \bibinfo{year}{2023}\natexlab{}.
\newblock \showarticletitle{Gpt-4 is too smart to be safe: Stealthy chat with llms via cipher}.
\newblock \bibinfo{journal}{\emph{arXiv preprint arXiv:2308.06463}} (\bibinfo{year}{2023}).
\newblock


\bibitem[Zhuo et~al\mbox{.}(2023)]%
        {zhuo2023exploring}
\bibfield{author}{\bibinfo{person}{Terry~Yue Zhuo}, \bibinfo{person}{Yujin Huang}, \bibinfo{person}{Chunyang Chen}, {and} \bibinfo{person}{Zhenchang Xing}.} \bibinfo{year}{2023}\natexlab{}.
\newblock \showarticletitle{Exploring ai ethics of chatgpt: A diagnostic analysis}.
\newblock \bibinfo{journal}{\emph{arXiv preprint arXiv:2301.12867}} (\bibinfo{year}{2023}).
\newblock


\bibitem[Zou et~al\mbox{.}(2023)]%
        {zou2023universal}
\bibfield{author}{\bibinfo{person}{Andy Zou}, \bibinfo{person}{Zifan Wang}, \bibinfo{person}{J~Zico Kolter}, {and} \bibinfo{person}{Matt Fredrikson}.} \bibinfo{year}{2023}\natexlab{}.
\newblock \showarticletitle{Universal and transferable adversarial attacks on aligned language models}.
\newblock \bibinfo{journal}{\emph{arXiv preprint arXiv:2307.15043}} (\bibinfo{year}{2023}).
\newblock


\end{thebibliography}

\end{document}